\documentclass[12pt]{article}
\usepackage{graphicx}
\usepackage{latexsym}
\begin{document}
\renewcommand{\theequation}{\thesection.\arabic{equation}}
\newcommand{\be}{\begin{equation}}
\newcommand{\ee}{\end{equation}}
\newcommand{\beqy}{\begin{eqnarray}}
\newcommand{\eeqy}{\end{eqnarray}}
\newcommand{\p}{\partial}
\newcommand{\hp}{\widehat{\p}}
\newcommand{\ov}{\overline}
\newcommand{\da}{^{\dagger}}
\newcommand{\w}{\wedge}
\newcommand{\st}{\stackrel}
\newcommand{\mb}{\mbox}
\newcommand{\mx}{\mbox}
\newcommand{\mt}{\mathtt}
\newcommand{\dt}{\mathtt{d}}
\newcommand{\al}{\alpha}
\newcommand{\bb}{\beta}
\newcommand{\ga}{\gamma}
\newcommand{\te}{\theta}
\newcommand{\Te}{\Theta}
\newcommand{\de}{\delta}
\newcommand{\et}{\tilde{e}}
\newcommand{\ze}{\zeta}
\newcommand{\s}{\sigma}
\newcommand{\e}{\epsilon}
\newcommand{\om}{\omega}
\newcommand{\Om}{\Omega}
\newcommand{\la}{\lambda}
\newcommand{\La}{\Lambda}
\newcommand{\n}{\nabla}
\newcommand{\hn}{\widehat{\nabla}}
\newcommand{\hph}{\widehat{\phi}}
\newcommand{\ah}{\widehat{a}}
\newcommand{\bh}{\widehat{b}}
\newcommand{\ch}{\widehat{c}}
\newcommand{\ddh}{\widehat{d}}
\newcommand{\eh}{\widehat{e}}
\newcommand{\gh}{\widehat{g}}
\newcommand{\ph}{\widehat{p}}
\newcommand{\qh}{\widehat{q}}
\newcommand{\mh}{\widehat{m}}
\newcommand{\nh}{\widehat{n}}
\newcommand{\Dh}{\widehat{D}}
\newcommand{\stu}{\st{\textvisiblespace}}
\newcommand{\au}{\stu{a}}
\newcommand{\bu}{\stu{b}}
\newcommand{\cu}{\stu{c}}
\newcommand{\du}{\stu{d}}
\newcommand{\eu}{\stu{e}}
\newcommand{\mmu}{\stu{m}}
\newcommand{\nnu}{\stu{n}}
\newcommand{\pu}{\stu{p}}
\newcommand{\Du}{\stu{D}}
\newcommand{\sto}{\st{\circ}}
\newcommand{\as}{\st{\circ}{a}}
\newcommand{\bs}{\st{\circ}{b}}
\newcommand{\cs}{\st{\circ}{c}}
\newcommand{\ds}{\st{\circ}{d}}
\newcommand{\es}{\st{\circ}{e}}
\newcommand{\ms}{\st{\circ}{m}}
\newcommand{\ns}{\st{\circ}{n}}
\newcommand{\ps}{\st{\circ}{p}}
\newcommand{\Ds}{\st{\circ}{D}}
\newcommand{\sts}{\st{s}}
\newcommand{\sth}{\st{\heartsuit}}
\newcommand{\stp}{\st{\perp}}
\newcommand{\std}{\st{\diamondsuit}}
\newcommand{\ad}{\st{s}{a}}
\newcommand{\bd}{\st{s}{b}}
\newcommand{\cd}{\st{s}{c}}
\newcommand{\gd}{\st{s}{g}}
\newcommand{\dd}{\st{s}{d}}
\newcommand{\Dd}{\st{s}{D}}
\newcommand{\ed}{\st{s}{e}}
\newcommand{\fd}{\st{s}{f}}
\newcommand{\zd}{\st{s}{\xi}}
\newcommand{\md}{\st{s}{m}}
\newcommand{\nd}{\st{s}{n}}
\newcommand{\stc}{\st{c}}
\newcommand{\az}{\st{c}{a}}
\newcommand{\bz}{\st{c}{b}}
\newcommand{\cz}{\st{c}{c}}
\newcommand{\dz}{\st{c}{d}}
\newcommand{\Dz}{\st{c}{D}}
\newcommand{\ez}{\st{c}{e}}
\newcommand{\fz}{\st{c}{f}}
\newcommand{\nz}{\st{c}{n}}
\newcommand{\mz}{\st{c}{m}}
\newcommand{\tb}{\overline{\theta}}
\newcommand{\ti}{\widetilde}

\newcommand{\2}{\frac{1}{2}}
\newcommand{\3}{\frac{1}{3}}
\newcommand{\4}{\frac{1}{4}}
\newcommand{\8}{\frac{1}{8}}
\newcommand{\6}{\frac{1}{16}}

\newcommand{\ra}{\rightarrow}
\newcommand{\Ra}{\Rightarrow}
\newcommand{\im}{\Longleftrightarrow}
\newcommand{\hs}{\hspace{5mm}}
\newcommand{\x}{\star}
\newcommand{\Delt}{\p^{\star}}
\newcommand{\vs}{\vspace{5mm}\\}

\thispagestyle{empty}
{\bf 24th July, 2004} 

\vspace{1cm}
\begin{center}{\Large{\bf COSMOLOGY FROM MODULI DYNAMICS }}\\
\vspace{1cm}
{\large{\bf Tirthabir Biswas\footnote{tirtho@hep.physics.mcgill.ca} and Prashanth Jaikumar\footnote{jaikumar@hep.physics.mcgill.ca} }}\\
\vspace{5mm}
{\small Centre for High Energy Physics\\
 Physics Department, McGill University\\
3600, University Street, Montreal H3A 2T8, Qu\'{e}bec, Canada\\}
\end{center}

\begin{abstract}
We investigate moduli field dynamics in supergravity/M-theory like set ups where we turn on fluxes along some or all of the extra dimensions. As has been argued in the context of string theory, we observe that the fluxes tend to stabilize the squashing (or shape) modes. Generically we find that at late times the shape is frozen while the radion evolves as a quintessence field. At earlier times we have a phase of radiation domination where both the volume and the shape moduli are slowly evolving.  However, depending on the initial conditions and the parameters of the theory, like the value of the fluxes, curvature of the internal manifold and so on,  the dynamics of the internal manifold  can be richer  with interesting cosmological consequences, including inflation. 
\end{abstract}

\newpage
\setcounter{page}{1}

\section{{\bf   INTRODUCTION}}

Unified Gravitational theories, dating back to Kaluza-Klein theory till the most modern String/M-theory, all seem to  require the existence of extra dimensions. Understanding their nature is therefore   of key interest. Is the ``internal manifold'' constituted by the extra dimensions stable, i.e. frozen in time, or is it evolving slowly? What are the problems and advantages associated with either of these scenarios? Even if the internal manifold is stable now, could it have had an earlier dynamic phase? In this paper we try to address some of these issues along with the cosmological and particle physics implications of these different scenarios.

 Recently in the context of String theory it has been argued that by turning on the fluxes one can fix the complex moduli  \cite{kachru}, which loosely speaking corresponds to the shape(s) of the internal manifold. One of the motivations for this paper is to study this mechanism of stabilizing the shape moduli explicitly in the context of supergravity (SUGRA)/M-theory type compactifications \cite{kaluza} using fluxes along some or all of the extra dimensions. Hence we start with a general bosonic\footnote{To study the moduli dynamics the fermionic fields can be set to zero.} SUGRA type action containing the usual Einstein scalar curvature term and various $n$-form fluxes. To keep the analysis as general as possible we also include a higher dimensional cosmological constant. For moduli fields, we choose to study the size and the shape or equivalently the volume and the squashing mode (preserving a subgroup, say  $H$, of the total isometry group $G$ of the internal manifold) which has been previously shown to have a consistent dimensional reduction \cite{duff,t2}. Such consistency arguments are known  to hold not only for pure gravitional action but also when  fluxes are turned on along all of the extra dimensions. Here we first show that it is also possible to ``consistently'' turn on fluxes only along  a  submanifold of the full internal manifold which preserves the  subgroup $H$ of isometries. We find, perhaps not surprisingly, that these extra fluxes have the effect of stabilizing the shape or the squashing mode in congruence with \cite{kachru}.
 
 Although in general the shape  is stabilized, the radion or the volume which is analogous to the Kahler moduli is not\footnote{For  attempts in this regard see \cite{kachru2}. This problem of stabilizing the volume or radion is also closely related to the issue of  dilaton stabilazation.}; it seems to have a runaway exponential potential. The advantage of having such a potential   is that the radion  can now act as a quintessence field \cite{ratra}, with the potential energy becoming vanishingly small as the volume rolls along the exponential. Such a potential can then naturally explain the exponential hierarchy between the currently observed cosmological energy density $<\la>$ and the (reduced) Planck mass $M_{\mt{p}}$ \cite{cosmos}
\be
<\la>=10^{-120}M_{\mt{p}}^4\ ,\label{cosmos}
\ee  
On the other hand if the radion is stabilized at a potential minima, then one has to typically invoke fine tuning to justify why the potential energy vanishes (or is very small (\ref{cosmos})) at the minima\footnote{As to why the quantum corrections to the vacuum energy are also as small as (\ref{cosmos}) is a much more difficult question to answer. One should mention that in brane world scenarios, specially in the six dimensional version \cite{cliff}, there is reason to believe that the quantum corrections may be of the right size.}. The flip side of having a runaway radion is that it typically couples  to matter-radiation. This first of all makes several coupling constants like the fine structure and Newton's constant time dependent (as they  depend on the volume which is now evolving in time) and there are strong constraints \cite{alpha,review} on the same. Second, since the quintessence field is effectively massless, it mediates a fifth force between matter/radiation  leading to violation of the equivalence principle which has been tested to high precision, leading to strong bounds \cite{review}. Is it  possible to avoid these bounds? 

The effective potential that we obtain for the moduli turns out to be a sum of six exponentials with complicated but rich features. In particular what is really interesting is that we find,  for certain values of group theoretic parameters and fluxes there are late time attractor solutions where a linear combination of the radion and the shape, rather than just the shape, is stabilized! Now note that radiation/matter will typically couple to a linear combination of the radion and the shape field. Thus if the combination that is stabilized coincides with (or is very close to)  the linear combination that couples to matter/radiation then the fifth force constraints will be easier to satisfy. In our case we find that although the introduction of the extra flux allows us to stabilize a linear combination which is close to the ``coupling'' linear combination, it is not close enough to avoid all the bounds. However, the results are encouraging and suggest that it may be possible to obtain the right linear combination by turning on fluxes in other directions or by accounting for potentials coming from brane-gas \cite{vafa} etc. Also, recently discovered chameleon effects of the scalar moduli fields \cite{justin} may play a role in ameliorating these constraints. Both the linear-combination and chameleon effects also have a tendency to moderate the constraints coming from time variation of physical constants. These latter type of  constraints however can also be avoided in another way. Since the kinetic energy of the quintessence field is related to the potential energy or the cosmological constant through the equation of state parameter, one can check that  
(\ref{cosmos}) implies
\be
 \dot{Q}\approx 10^{-13}-10^{-14} yr^{-1}
\label{0qdot}
\ee
where $Q$ is the quintessence field, in general a linear combination of the moduli. 
The typical age of the universe being $10^{10}$ years, (\ref{0qdot}) suggests a very small variation of $Q$ and hence of the physical constants (including the cosmological constant), well within the observational limits since big bang nucleosynthesis (BBN). This presents an alternative scenario where most of the evolution of the scalar fields occurs before BBN, while since then all the scalar fields are effectively frozen \cite{frozen}! We discuss in details some of the  consequences and observational bounds of having ``rolling'' moduli fields.   

However this is not the only possible scenario.  Depending on the various parameters our potential can have one or two minima. The first one, which is always present, corresponds to the symmetric unsquashed state where the internal manifold preserves all the $G$ isometries, while the second one can arise due the presence of the extra flux and its depth depends on the value of the flux. Thus it is possible that the depth of this minimum is so large that once our universe gets stuck it remains there. This leads to a different  scenario where all the moduli are fixed! Of course in this case the fine tuning issues of the classical or bare cosmological constant remains unsolved. What could be really interesting in these situations is that even though the internal manifold is now frozen it could  have been dynamic before, with possible application to inflation \cite{guth,linde}. Imagine the universe starting out at the ``shallow'' symmetric minimum and then making a transition, either to the second minimum or towards the ``runaway behaviour''. As explained in \cite{t1} this would correspond to a gauge symmetry breaking mechanism, $G\ra H$, in the effective four dimensional world\footnote{In the braneworld interpretation squashing of the internal manifold will not be related to any gauge symmetry breaking in the brane because the  bulk gauge fields are different from the ones living on the brane.}; typically the mass of the broken gauge bosons corresponds to the scale of Grand Unified theories (GUTs) \cite{t1,t2}. During this phase transition, if the inflationary slow roll conditions \cite{linde} are satisfied around the top of the potential barrier, then our universe will experience an accelerated growth or inflation. Let us now briefly focus on some of the important details of the analysis and results. 

We  developed general and physically intuitive techniques to study the dynamics governed by a sum of exponential potentials where the exponents are arbitrary linear combinations of two fields\footnote{See \cite{riet} for a recent and more formal analysis of critical points in such systems.}. These techniques are based on a rich body of literature \cite{liddle,anupam,wohlfarth} on exponential potentials and in priciple could be extended to more than two fields. Also, recently a more formal analysis of critical points have been performed The analysis relies on finding the asymptotic behaviour of the potential as one moves towards infinity  along any direction on the ``moduli-plane''. Since we have only exponentials, ultimately only a single exponent dominates;  the potential either grows  exponentially or falls to zero. In this way one can map the entire moduli-plane into regions where each exponent dominates (see fig.\ref{figps3}). Since the dynamics of a single exponential is exactly solvable \cite{liddle} one can construct qualitative trajectories of the ``particle'' as it passes through different regions dominated by different exponents. What is remarkable is that we find there are two ultimate late time attractor states of the particle (one when the shape, and the other when a linear combination of the shape and the radion is stabilized) and both of these solutions correspond to an effective exponent which is less than $\sqrt{2}$ signalling a phase of acceleration or quintessence. On the other hand all other exponents are larger than $\sqrt{2}$ corresponding to radiation-matter domination\footnote{For radiation domination one really needs the exponent to be larger than $\sqrt{3}$ and this is also acheived for several values of the various dimensions in our model.}! 

Let us try to  summarize the dynamics. We investigate three possible initial conditions depending upon which  cosmology may vary: (I-1) the universe  starts at the symmetric minimum, (I-2) the universe starts away from minimum  with large ($\gg M_p$) potential energies and as it rolls down it is able to completely avoid the symmetric minimum, (I-3) although the universe starts away from the minimum, there are mechanisms which depending upon the parameters can suck the particle close to the minimum after which the dynamics is similar to that of (I-1). We find that there are three possible late time states: (L-1) quintessence phase when the shape is stable, (L-2), another quintessence phase where a combination of the shape and the volume is frozen, and (L-3) the universe is stabilized at the second minimum. The intermediate dynamics is typically non-accelerating (radiation-matter dominated) except a transitional phase where the particle comes out of the symmetric minimum in (I-1) and (I-3), and this  can be accompanied by inflation. The aim of this paper is not to come up with ``The Scenario'' where everything works but to rather motivate how different phases of cosmology: inflation, radiation-matter domination and quintessence can arise from moduli dynamics of SUGRA/String theory. Also, the paper develops analytical techniques to analyze complicated potentials involving exponentials that is expected in unified gravitational theories. These should be useful in further research to evolve a realistic cosmological history. Finally, the reader is warned that some of the analysis may appear too technical and detailed but considering the generic nature of the potential for SUGRA models the authors feel that it is important to study the dynamics in details. 

The paper is organized as follows: In section 1, we introduce our higher dimensional SUGRA type model with a brief review of pertinent group theoretical concepts. In section  2, we perform a consistent dimensional reduction including the fluxes. In section 3, the effective action and couplings to radiation is derived. In section 4, techniques to analyse dynamics with exponential potentials are developed and applied to understand the general features of the cosmological evolution of the moduli. In section 5, we try to take steps towards realistic cosmological scenarios with emphasis on inflation and quintessence. In particular, we study the various observational constraints on inflationary and quintessence cosmology. Finally, we summarize our results and indicate the challenges that need to be overcome. 
\section{{\bf  OUR MODEL }}
As in \cite{t1,t2} we consider our universe to be a semi-direct product, $M_{D+1}\otimes G$, where $M_{D+1}$ is the $D+1$-dimensional observational universe and $G$, a Lie group manifold\footnote{Although for simplicity we choose to work with a group manifold, most of the analysis depends only on the various dimensions entering into the problem and hence should hold at least qualitatively  for more general homogeneous spaces like the coset spaces.}, serves as the Kaluza-Klein internal space \cite{kerner}. Let us first review the Lie group geometry in brief. 
\vspace{5mm}
\\
{\bf Geometry of Lie groups:}  A Lie group element $g$ can be parameterised as 
\begin{equation}
g=exp(\chi^{\as}(y^{\ms})T_{\as})\ \in\ G
\end{equation}
where $T_{\as}\ \in\ \cal{G}$, the Lie algebra corresponding to the Lie group $G$ and $\chi^{\as}(y^{\ms})$ are some given functions of the coordinates $y^{\ms}$ charting the Lie group manifold. The Lie group generators $T_{\as}$ satisfy the usual commutation relations:
\begin{equation}
[T_{\as},T_{\bs}]=C_{\as\bs}{}^{\cs}T_{\cs}
\end{equation}
where $C_{\as\bs}{}^{\cs}$ are the structure constants of the Lie group. With each of the generators $T_{\as}$, one can associate a left and a right invariant vector field $e_{\as}$ and $\et_{\as}$ respectively.

We are interested in metrics which are at least invariant under the action of the right invariant vector fields. The components of such metrics are constants if we choose the left invariant vector fields $e_{\as}$, defined by 
$$
e_{\as}\equiv e_{\as}{}^{\ms}\p_{\ms};\ e_{\as}{}^{\ms}\equiv (e_{\ms}{}^{\as})^{-1}
$$
and
\begin{equation}
g^{-1}\p_{\ms}g=e_{\ms}{}^{\as}T_{\as}\ , 
\end{equation}
as the local veilbien basis. A special case of the left invariant metric is the bivariant which is invariant under both the left and the right invariant vectors. The Killing metric given by
\begin{equation}
g^K_{\as\bs}=-C_{\as\cs}{}^{\ds}C_{\bs\ds}{}^{\cs}
\end{equation}
is an example of such a metric. Further, the Killing metric satisfies Einstein's field equations
\begin{equation}
R_{\as\bs}=\sto{\la}g_{\as\bs}
\label{eq:4einstein}
\end{equation}
and hence is consistent with its usual identification as Kaluza-Klein vacuum,
the constant $\sto{\la}$ being referred to as the internal curvature. Contrary to this picture of an  internal manifold frozen in its maximally symmetric Killing metric, we treat it as  dynamic. In particular, it is interesting  to study whether the manifold makes a transition from the $G_L\otimes G_R$ Killing metric to a  $G_L\otimes H_R$-invariant ``squashed'' metric, thereby effecting a gauge symmetry breaking from $G_R\ra H_R$ in four dimensions~\cite{t1}. The metric in this case looks like
\begin{equation}
g^S_{\as\bs}=\Psi^2\left( \begin{array}{cc}
g^K_{\az\bz} & 0\\
0& \Te^2g^K_{\ad\bd}
\end{array} \right)
\label{eq:wmetric}
\end{equation}
Here and from now on we refer group quantities by a circle ($\circ$) while that of the  {\em C}oset space $G/H$ and the {\em S}ubgroup $H$ with ($c$) and  ($s$) respectively. Sometimes we may omit the symbols when it is self-evident. We will also assume the group $G$ to be semi-simple and the coset decomposition to be reductive 
\be
C_{\az\bd}{}^{\cd}=0
\ee 
In (\ref{eq:wmetric}) $\Psi^2$ corresponds to the ``size'' of the internal manifold and $\Te^2$ is the ``squashing'' parameter. Note, that we are in a broken phase when $\Te^2\neq 1$ and thus, as noted in \cite{t1,t2}, symmetry breaking can take place via two kinds of transition: (a) $\Te^2$ can make a transition from a symmetric vacuum $\Te^2= 1$ to a non-symmetric vacuum at $\Te^2= \Te_0^2\neq 1$ indicating a Higg's like mechanism, and (b) $\Te^2$ can keep evolving very slowly towards infinity much like a quintessence field, effecting what we call a quintessential transition!
\vspace{5mm}\\
{\bf Action and the  Ansatz:} As explained in the introduction, our aim is to try to connect the cosmology and particle physics that we observe in our four dimensional world with  unified higher dimensional theories, like Supergravity/M-theory. Typically the bosonic sector of such theories consists of the metric, some form fields and occassionally the dilaton. Thus, our starting point is the higher dimensional action given by
\begin{equation}
\hat{S}=\frac{1}{16\pi G_{\Dh}}\int dx^{\Dh}\sqrt{-g}\{\hat{R}-\2\p_{\mh}\phi\p^{\mh}\phi-\2\sum_I \frac{1}{n_I!}e^{a_I\phi}F^2_{n_I}-2\hat{\La}\}
\label{eq:sugra}
\end{equation}
where $\phi$ is the dilaton field and the field strengths $F_{n_I}$'s are  $n_I$  forms, $I=1\dots M$ and we have also included a cosmological constant term.

To obtain an effective four-dimensional theory one has to perform a consistent  dimensional reduction \cite{duff}. One then ends up with Einstein's theory of gravity coupled to several ``reduced'' form fields and a bunch of scalars corresponding to the moduli of the internal manifold. Since one can usually interpret the dilaton as the radial moduli corresponding to a one dimensional circular compactification, we won't include it seperately in our discussion from here on. Also, although both in the action and in the field strengths one can have Chern-Simons-like terms, they can be ignored for the  ansatz that we make. 

 Since we want to study the symmetry breaking mechanism we have to include the squashing mode, while the size or the breathing mode has to be included for consistency. Including the shape or the squashing field has the additional advantage of being able to produce non-trivial curvature potentials\footnote{ In a cosmological context such potentials were first  studied for specific coset spaces in \cite{pope} focussing on cosmological instanton and domain wall solutions. In \cite{t2} squashing in arbitrary compact group manifolds was  discussed while in \cite{ohta} cosmological solutions, where the internal manifold is a product space including having a hyperbolic geometry, has been studied. More recently in \cite{transient} three dimensional group manifolds, both compact and noncompact,  have been studied in details.}  which can enrich the cosmology that one obtains. In the Kaluza-Klein reduction scheme we thus have to upgrade $\Psi\ra \Psi(x)$ and $\Te\ra \Te(x)$. We will denote the coordinates charting the observable universe $M_{D+1}$ by $x^m$ while $x^{\mh}$ will be used to collectively  denote $\{x^m,y^{\ms}\}$. As is perhaps evident by now, we are  using ``hatted'', $(\ \widehat{ }\ )$,  quantities to refer to objects corresponding to the full higher dimensional manifold. 

Although an expression of the metric of the form (\ref{eq:wmetric}) is physically clarifying, technically it is more convenient to include the scalars in the vielbein. We choose to  parameterise the group element as 
\be
g=exp(\chi^{\az}(y^{\mz})T_{\az}) exp(\chi^{\ad}(z^{\md})T_{\ad})
\label{1coord}
\ee
 The ansatz for the full higher dimensional vielbein is then given by
\begin{equation}
\eh_{\mh}{}^{\ah}=\left( \begin{array}{ccc}
e_m{}^{a}(x) & 0&0\\
0 &\Psi(x)\es_{\mz}{}^{\az}(y,z)&\Psi(x)\Te(x)\es_{\mz}{}^{\ad}(y,z)\\
0&0& \Psi(x)\Te(x)\es_{\md}{}^{\ad}(z)
\end{array} \right)
\label{eq:4s-inv}
\ee
and
\be
\eh_{\ah}{}^{\mh}=\left( \begin{array}{ccc}
e_{a}{}^{m}(x) & 0&0\\
0 &\ \Psi^{-1}(x)\es_{\az}{}^{\mz}(y,z)&\Psi^{-1}(x)\es_{\az}{}^{\md}(y,z)\\
0&0& \Psi^{-1}(x)\Te^{-1}(x)\es_{\ad}{}^{\md}(z)
\end{array} \right)
\label{eq:4s-vielbein}
\end{equation}
The flat-metric is then just a constant
\begin{equation}
\widehat{g}_{\ah\bh}=\left( \begin{array}{cc}
\eta_{ab} & 0\\
0 & g^K_{\as\bs}
\end{array} \right)
\label{eq:4s-metric}
\end{equation}
We did not include the vectors in the ansatz (\ref{eq:4s-inv}-\ref{eq:4s-metric}) because we are only interested in the vacuum dynamics and the vectors appear as fluctuations around the vacuum metric.

For the fluxes, several different choices of ansatz are possible depending upon the degree of the form field. Note that we are also not interested in the dynamics (or fluctuations) of the form fields but rather the condensate or vacuum expectation value which typically depends on the moduli fields once one solves the field equations that follow from (\ref{eq:sugra}) \cite{pope}
\begin{equation}
\frac{1}{(n-1)!}\frac{1}{\sqrt{-\gh}}\p_{\mh}(\sqrt{-\gh}F^{\mh\ph_2\dots\ph_I})=0
\label{eq:gauge}
\end{equation}
 and the Bianchi identity
\be
\ddh F=0
\label{eq:bianchi}
\ee
Thus the $F^2$ term in the action gives us an additional potential term for scalars.

 Now, depending upon the exact internal manifold and the non-trivial forms that it can support one can have non-zero fluxes in different directions. However, in our model some of the choices appear naturally:\\
1. Flux along $G$:  
$$F^{\Ds}_{\ms_1\dots\ms_{\Ds}}\sim \e_{\ms_1\dots\ms_{\Ds}}$$
2. Flux along $H$:
$$F^{\Dd}_{\md_1\dots\md_{\Dd}}\sim \e_{\md_1\dots\ms_{\Dd}}$$

Equipped with the ansatz we can now proceed to perform a consistent dimensional reduction.
\vspace{5mm}
\\ 
\setcounter{equation}{0}
\section{ {\bf CONSISTENT DIMENSIONAL REDUCTION }}
{\bf Effective Action from Gravity:} First let us look at the Einstein-Hilbert term in the higher dimensional action (\ref{eq:sugra}).  To compute the scalar curvature corresponding to the ansatz (\ref{eq:4s-inv}-\ref{eq:4s-metric}) it is convenient to use differential forms. One starts with the basis 1-forms
\be
\widehat{\om}^{\ah}=dx^{\mh}e_{\mh}{}^{\ah}
\ee
For (\ref{eq:4s-inv}) the 1-forms are given by
$$\widehat{\om}^{a}=\om^a$$
$$\widehat{\om}^{\az}=\Psi\om^{\az}$$
\be
\widehat{\om}^{\ad}=\Psi\Te\om^{\ad}
\ee
One then proceeds to compute the Ricci tensor via the spin connections and the curvature 2-forms in the following manner: The spin connections $\widehat{\om}^{\ah}{}_{\bh}$  are uniquely defined via
\be
\ddh \widehat{\om}^{\ah}+\widehat{\om}^{\ah}{}_{\bh}\w\widehat{\om}^{\bh}=0
\label{eq:4connection}
\ee
The curvature 2-forms can be computed from the spin connections:
\be
\widehat{\cal{R}}^{\ah}{}_{\bh}=\ddh \widehat{\om}^{\ah}{}_{\bh}+\widehat{\om}^{\ah}{}_{\ch}\w\widehat{\om}^{\ch}{}_{\bh}
\ee
The coefficients of the Riemannian tensor can now be read off  from the curvature 2-forms
\be
\widehat{\cal{R}}^{\ah}{}_{\bh}=\widehat{R}^{\ah}{}_{\bh|\ch\ddh|}\widehat{\om}^{\ch}\w\widehat{\om}^{\ddh}
\ee
Without going into the details here, \cite{t2} we enumerate the Ricci tensor obtained from the usual contraction of the Remannian tensor
\be
\widehat{R}_{\bh \ddh}=\widehat{R}^{\ah}{}_{\bh \ah\ddh}
\ee
After some simplifications one obtains
$$\widehat{R}_{a b}=R_{a b}-\Ds\Psi^{-1}\n_b(e_a\Psi)-\Dd\Te^{-1}\n_b(e_a\Te)-\Dd\Psi^{-1}\Te^{-1}e_{(a}\Psi e_{b)}\Te$$
$$\widehat{R}_{\az \bz}=g_{\az \bz}[-\{\Psi^{-1}\Box\Psi+(\Ds-1)\Psi^{-2}(\p\Psi)^2+\Dd\Psi^{-1}\Te^{-1}\p_{a}\Psi \p^{a}\Te\} 
$$
$$+\2\Psi^{-2}\left\{1-\frac{k'}{2}-\frac{1-k'}{2}\Te^2\right\}]$$
$$
\widehat{R}_{\ad \bd}=g_{\ad \bd}[-\{\Psi^{-1}\Box\Psi+\Te^{-1}\Box\Te+(\Ds-1)\Psi^{-2}(\p\Psi)^2+(\Dd-1)\Te^{-2}(\p\Te)^2$$
\be
+(\Dd+\Ds)\Psi^{-1}\Te^{-1}\p_{a}\Psi \p^{a}\Te\}+\4\Psi^{-2}\{\Te^2(1-k)+\Te^{-2}k\}]
\label{eq:4ricci}
\ee
Here we have  assumed\footnote{These assumptions are not true for an arbitrary subgroup $H$ of $G$.} that
\be
\sts{g}_{\ad \bd}^K=k\sto{g}_{\ad \bd}^K\im C_{\ad\cd}{}^{\dd}C_{\bd\dd}{}^{\cd}=kC_{\ad\cs}{}^{\ds}C_{\bd\ds}{}^{\cs}
\ee
and 
\be
 C_{\az\cz}{}^{\dz}C_{\bz\dz}{}^{\cz}=k'C_{\az\cs}{}^{\ds}C_{\bz\ds}{}^{\cs}
\ee
Note $k$ and $k'$ are not independent group theoretical quantities but are related by
\be
1-k'=\frac{2\Ds}{\Dz}(1-k)
\ee
 For a symmetric coset decomposition, i.e. $C_{\az\cz}{}^{\dz}=0$ we have
\be
k'=0\Ra k=1-\frac{\Dz}{2\Dd}
\ee
Another interesting case is when $H=U(1)$ or a product of $U(1)$'s. Then 
\be
k=0\Ra k'=1-\frac{2\Dd}{\Dz}
\ee

We are ready to compute the scalar curvature that we need to compute the Einstein-Hilbert term in the action (\ref{eq:sugra}):
\be
\widehat{S}_{\widehat{D}}=\int dx^{\widehat{D}}\ \eh^{-1}\widehat{R}
\label{eq:4action}
\ee
Now
\be
\widehat{R}=g^{\ah \bh}\widehat{R}_{\ah \bh}=g^{a b}\widehat{R}_{a b}+g^{\az \bz}\widehat{R}_{\az \bz}+g^{\ad \bd}\widehat{R}_{\ad \bd}
\ee
or,
$$\widehat{R}=R-\left[2\Ds\frac{\Box\Psi}{\Psi}+2\Dd\frac{\Box\Te}{\Te}+\Ds(\Ds-1)\frac{(\p\Psi)^2}{\Psi^2}+\Dd(\Dd-1)\frac{(\p\Te)^2}{\Te^{2}}+2\Dd(\Ds+1)\frac{\p_{a}\Psi \p^{a}\Te}{\Psi\Te}\right]$$
\be
+\frac{1}{4\Psi^2}\left[  \Dz+2\Dd(1-k)-\Dd(1-k)\Te^2+ k\Dd \frac{1}{\Te^2} \right]
\label{eq:4curvature}
\ee

Since $\widehat{R}$ is independent of the group coordinates one can perform the integration over the group in the action (\ref{eq:4action}) which essentially just yields a volume factor
\be
V_G=\int d^{\Ds}y \es^{-1}
\ee
 Thus we have our effective $D+1$-dimensional action
\be
S_{\mt{grav}}=\frac{V_G}{16\pi \hat{G}}\int e^{-1}\Psi^{\stackrel{\circ}{D}}\Te^{\Dd}\widehat{R}=\frac{1}{16\pi G}\int e^{-1}\Psi^{\stackrel{\circ}{D}}\Te^{\Dd}\widehat{R}=\frac{M_p^2}{2}\int e^{-1}\Psi^{\stackrel{\circ}{D}}\Te^{\Dd}\widehat{R}
\ee
where we have identified
\be
\frac{M_p^2}{2}\equiv\frac{1}{16\pi G_{D+1}}\equiv \frac{V_G}{16\pi G_{\Dh}}\,
\ee
$M_p$ being the four dimensional reduced Planck mass, $10^{18}$ Gev. For most part of the analysis we work in natural units where $M_p=1$.
It is useful to perform some integration by parts. The simplified action looks like
\be
S_{\mt{grav}}=\frac{1}{2}\int dx^{D+1}\ e^{-1}\Psi^{\Ds}\Te^{\Dd}[R-K+V]
\ee
where we have defined the Kinetic and Potential like terms\footnote{Note that the coefficients in front of the kinetic terms only depend on the various dimensions  and is therefore completely general for any internal manifold. The coefficients of the potential terms however will vary as one passes from   group manifolds to more general manifolds, although the functional dependence on $\Te$ and $\Psi$ remains the same.} for the scalar fields as
\be
K=-\left[\Ds(\Ds-1)\frac{(\p\Psi)^2}{\Psi^2}+\Dd(\Dd-1)\frac{(\p\Te)^2}{\Te^{2}}+2\Dd(\Ds-1)\frac{\p_{a}\Psi \p^{a}\Te}{\Psi\Te}\right]
\label{2kinetic}
\ee
and
\be
V=\frac{1}{4\Psi^2}\left[ -( \Dz+2\Dd(1-k))+\Dd(1-k)\Te^2- k\Dd \frac{1}{\Te^2} \right]
\ee

One can also compute the effective action term corresponding to the   higher dimensional cosmological constant
\be
S_{\mt{cos}}=-2\hat{\La} \frac{1}{2}\int dx^{D+1}\ e^{-1}\Psi^{\Ds}\Te^{\Dd}
\label{eq:4c-action}
\ee
\vspace{5mm}
\\ 
{\bf The Fluxes:} Next, let us  solve the Bianchi identity (\ref{eq:bianchi}) and the field equations (\ref{eq:gauge}) for the field strengths. For simplicity we  assume that only one of the $n$-forms discussed above is turned on.\\
1. Flux along $G$ \cite{kaluza}: The Bianchi identity essentially tells us
\be
F^{\Ds}_{\ms_1\dots\ms_{\Ds}}=\,\cs(y)\e_{\ms_1\dots\ms_{\Ds}}
\label{2coF}
\ee
Then
\be
F^{\Ds,\ms_1\dots \ms_{\Ds}}=(\hat{g}^{-1})_{\Ds}\cs(y)\e^{\ms_1\dots \ms_{\Ds}}=\frac{1}{\Psi^{2\Ds}\Te^{2\Dd}}\sto{g}^{-1}\es^{2}(y)\cs(y)\e^{\ms_1\dots \ms_{\Ds}}
\label{2contraF}
\ee
Plugging (\ref{2contraF}) in (\ref{eq:gauge}) we find
\be
\cs(y)=\,\cs.\es(y)^{-1}
\ee
with the flux energy reading
\be
F^2 =\,\cs^2(D+1)!\sto{g}^{-1}\frac{1}{\Psi^{2\Ds}\Te^{2\Dd}}
\ee
The effective action is then given by
\be
S_F=-\4 M_p^2\cs^2\int d^{D+1}x\ e^{-1}\frac{1}{\Psi^{\Ds}\Te^{\Dd}}
\ee
where we have absorbed $\sto{g}^{-1}$ in the definition of $\cs$. The above flux contribution is well known in literature but we now come to a more non-trivial flux contribution along the subgroup $H$  which as we shall see later, is consistent to turn on under certain conditions.\\
2. Flux along $H$: As usual the Bianchi identity implies
\be
F^{\Dd}_{\md_1\dots\md_{\Dd}}=\cd(z) \e_{\md_1\dots\ms_{\Dd}}
\ee
Then
$$F^{\Dd,\md_1\dots \md_{\Dd}}=(\hat{g}^{-1})_{\Dd}\cd(z)\e^{\md_1\dots \md_{\Dd}}=-\ed^{2}(z)\sts{g}^{-1}\cd(z)\e^{m_1\dots m_{D+1}}\frac{1}{\Psi^{2\Dd}\Te^{2\Dd}}$$
Now 
$$\sqrt{-\hat{g}}=e^{-1}(x)\ez^{-1}(y)\sqrt{\sto{g}}D(z)\ed^{-1}(z)\Psi^{\Ds}\Te^{\Dd}$$
where $D(z)$ is the determinant of the adjoint representation of $H$ in $G/H$, $D_{\az}{}^{\bz}(h(z))$ and $\ez(y)$ is the determinant of the vielbein along the coset space $G/H$. This is possible because in our choice of co-ordinate system (\ref{1coord}) the vielbein decomposes so that
$$\es_{\mz}{}^{\az}(y,z)=\ez_{\mz}{}^{\bz}(y)D_{\bz}{}^{\az}(h(z))$$
As we shall soon discover, for consistency one requires $D(z)$ to be a constant, $d$.
Plugging the metric and the field strength now in the field equations (\ref{eq:gauge}) one finds that it is satisfied if 
\be
\cd(z)=\cd.\ed^{-1}(z)
\ee
The vacuum flux energy reads
\be
F^2=\cd^2\Dd!\sts{g}^{-1}\frac{1}{\Psi^{2\Dd}\Te^{2\Dd}}
\ee
and  the effective action is then given by
\be
S_F=-\4 M_p^2\cd^2\int d^{D+1}x\ e^{-1}\frac{\Psi^{(\Ds-2\Dd)}}{\Te^{\Dd}}
\ee
where again we have redefined $\cd$ to absorb $\sts{g}^{-1}$.
\vs
{\bf Consistency of the Truncation:} 
In the previous subsection we obtained the dimensionally reduced field theoretic action for our model. It is important that we check the consistency of our ansatz \cite{duff}, i.e.  the solutions that one obtains by varying the effective action (\ref{eq:sugra}) should indeed be solutions of the full higher dimensional Einstein's equations. In \cite{t2} it was shown that the truncation is consistent when we only have the Einstein-Hilbert term. More specifically it was found that
\be
\left(\frac{1}{\Psi^{\Ds}\Te^{\Dd-1}}\frac{\de S_{\mt{grav}}}{\de \Te}\right)\widehat{g}_{\ad\bd}=- M_p^2e^{-1}\Dd\widehat{G}_{\ad\bd}
\label{2theta}
\ee
and
\be
\left(\frac{1}{\Psi^{\Ds-1}\Te^{\Dd}}\frac{\de S_{\mt{grav}}}{\de \Psi}-\frac{1}{\Psi^{\Ds}\Te^{\Dd-1}}\frac{\de S_{\mt{grav}}}{\de \Te}\right)\widehat{g}_{\az\bz}=-M_p^2e^{-1}\Dz\widehat{G}_{\az\bz}
\label{2psi}
\ee
Consistency of the ansatz (\ref{eq:4s-inv}-\ref{eq:4s-metric}) is now obvious\footnote{Since 
$\widehat{g}^{mn}=g^{mn}$,
i.e. there has been no field redefinition involving the four dimensional part of the metric it is clear that field equations corresponding to the variation of the four dimensional metric is automatically consistent.} from (\ref{2theta}) and (\ref{2psi}).

We now want to investigate whether the consistency is violated when we introduce the fluxes. To do that we need the field equations for the higher dimensional metric corresponding to the full higher dimensional action (\ref{eq:sugra}):
\be
G_{\ah\bh}-\2\sum_I \frac{1}{n_I!}(n_IF_{\ah\ch_2\dots\ch_{n_I}}F_{\bh}{}^{\ch_2\dots\ch_{n_I}}-\2 g_{\ah\bh}F^2_{n_I})+\hat{\La}g_{\ah\bh}=0
\ee
Now, as in the case with no flux, there is no field redefinition involved in the ordinary space-time component of the metric and hence we only need to worry about the field equations along the internal dimensions. Specializing to a single flux we thus have
\be
G_{\as\bs}-\2 \frac{1}{n!}(nF_{\as\ch_2\dots\ch_{n}}F_{\bs}{}^{\ch_2\dots\ch_{n}}-\2 g_{\as\bs}F^2_{n})+\hat{\La}g_{\as\bs}\equiv G_{\as\bs}-T^F_{\as\bs}-T^{\mt{cos}}_{\as\bs}=0
\label{eq:einstein}
\ee
In order for the consistency arguement to hold we need to show\footnote{It is easy to check that the cosmological term satisfies the analogous equations.}
\be
\left(\frac{1}{\Psi^{\Ds}\Te^{\Dd-1}}\frac{\de S_{\mt{flux}}}{\de \Te}\right)\widehat{g}_{\ad\bd}= M_p^2e^{-1}\Dd T_{\ad\bd}
\ee
and
\be
\left(\frac{1}{\Psi^{\Ds-1}\Te^{\Dd}}\frac{\de S_{\mt{flux}}}{\de \Psi}-\frac{1}{\Psi^{\Ds}\Te^{\Dd-1}}\frac{\de S_{\mt{flux}}}{\de \Te}\right)\widehat{g}_{\az\bz}= M_p^2e^{-1}\Dz T_{\az\bz}
\ee
Let us verify this case by case:\\
1. Flux along $G$:  It is easy to see
$$
\2 \frac{1}{\Ds!}\Ds F_{\as\ch_2\dots\ch_{\Ds}}F_{\bs}{}^{\ch_2\dots\ch_{\Ds}}=\2 \cs^2\sto{g}^{-1}\left(\frac{1}{\Psi^{2\Ds}\Te^{2\Dd}}\right)g_{\as\bs}$$
Then from (\ref{eq:einstein}) we have
$$T_{\as\bs}=\4 \cs^2\sto{g}^{-1}\left(\frac{1}{\Psi^{2\Ds}\Te^{2\Dd}}\right)g_{\as\bs}$$
Now let us compute the variation of the reduced action:
\be
\frac{\de S_{\mt{flux}}}{\de \Te}=\4 M_p^2e^{-1}\Dd \cs^2\sto{g}^{-1}\frac{1}{\Psi^{\Ds}\Te^{\Dd+1}}
\ee
and
\be
\frac{\de S_{\mt{flux}}}{\de \Psi}=\4 M_p^2e^{-1}\Ds \cs^2\sto{g}^{-1}\frac{1}{\Psi^{\Ds+1}\Te^{\Dd}}
\ee
Then
$$
\left(\frac{1}{\Psi^{\Ds}\Te^{\Dd-1}}\frac{\de S_{\mt{flux}}}{\de \Te}\right)\widehat{g}_{\ad\bd}=\4 M_p^2e^{-1}\Dd \cs^2\sto{g}^{-1}\frac{\widehat{g}_{\ad\bd}}{\Psi^{2\Ds}\Te^{2\Dd}}= M_p^2e^{-1}\Dd T_{\ad\bd}
$$
Similarly,
$$
\left(\frac{1}{\Psi^{\Ds-1}\Te^{\Dd}}\frac{\de S_{\mt{flux}}}{\de \Psi}-\frac{1}{\Psi^{\Ds}\Te^{\Dd-1}}\frac{\de S_{\mt{flux}}}{\de \Te}\right)\widehat{g}_{\az\bz}= \4 M_p^2e^{-1}(\Ds-\Dd) \cs^2\sto{g}^{-1}\frac{\widehat{g}_{\az\bz}}{\Psi^{2\Ds}\Te^{2\Dd}}=M_p^2e^{-1}\Dz T_{\az\bz}
$$
2. Flux along $H$: As in the earlier cases it is straight-forward to compute $T_{\as\bs}$. We find
\be
T_{\az\bz}=-\4 \cd^2\sto{g}^{-1}\left(\frac{1}{\Psi^{2\Dd}\Te^{2\Dd}}\right)g_{\az\bz}
\ee 
and 
\be
T_{\ad\bd}=\4 \cd^2\sto{g}^{-1}\left(\frac{1}{\Psi^{2\Dd}\Te^{2\Dd}}\right)g_{\ad\bd}
\ee 
The variations are given by
\be
\frac{\de S_{\mt{flux}}}{\de \Te}=\4 M_p^2e^{-1}\Dd \cd^2\sto{g}^{-1}\frac{\Psi^{\Ds-2\Dd}}{\Te^{\Dd+1}}
\ee
and
\be
\frac{\de S_{\mt{flux}}}{\de \Psi}=-\4 M_p^2e^{-1}(\Ds-2\Dd) \cd^2\sto{g}^{-1}\frac{\Psi^{\Ds-2\Dd-1}}{\Te^{\Dd}}
\ee
Then again we have
$$
\left(\frac{1}{\Psi^{\Ds}\Te^{\Dd-1}}\frac{\de S_{\mt{flux}}}{\de \Te}\right)\widehat{g}_{\ad\bd}=\4 M_p^2e^{-1}\Dd \cd^2\sto{g}^{-1}\frac{\widehat{g}_{\ad\bd}}{\Psi^{2\Ds}\Te^{2\Dd}}= M_p^2e^{-1}\Dd T_{\ad\bd}
$$
Similarly,
$$
\left(\frac{1}{\Psi^{\Ds-1}\Te^{\Dd}}\frac{\de S_{\mt{flux}}}{\de \Psi}-\frac{1}{\Psi^{\Ds}\Te^{\Dd-1}}\frac{\de S_{\mt{flux}}}{\de \Te}\right)\widehat{g}_{\az\bz}= -\4 M_p^2e^{-1}(\Ds-2\Dd+\Dd) \cd^2\sto{g}^{-1}\frac{\widehat{g}_{\az\bz}}{\Psi^{2\Ds}\Te^{2\Dd}}$$
$$=M_p^2e^{-1}\Dz T_{\az\bz}
$$
Thus we have shown that we can have a consistent dimensional reduction even with flux turned on along $G$ or $H$. Also, since in general the degree of the flux-form is different in the two different cases, the field equations and consequently the consistency does not interfere with each other, and one can imagine all the fluxes turned on simultaneously. 
\section{ {\bf EFFECTIVE ACTIONS AND COUPLINGS}}
In the previous section we performed a consistent dimensional reduction of a higher dimensional supergravity type action containing the pure Einstein-Hilbert term along with a cosmological constant and fluxes. To analyze the cosmological and particle physics aspect of the model, it is however convenient to perform conformal rescalings so that the Einstein-Hilbert term is canonical in four dimensions.
\vspace{5mm}
\\ 
{\bf The Conformal Transformations:}  Let us first redefine the moduli fields to bring the kinetic terms (\ref{2kinetic}) in the usual form:
\be
\Psi=e^{\psi};\mbox{ and }\Te=e^{\te}
\ee
In the new variables the gravitational action looks like
\be
S_{\mt{grav}}=\2 M_p^2\int dx^{D+1}\ e^{-1}e^{\Ds\psi+\Dd\te}[R-K+V_{\mt{grav}}]
\label{eq:4e-action}
\ee
where
\be
K=-\left[\Ds(\Ds-1)(\p\psi)^2+\Dd(\Dd-1)(\p\te)^2+2\Dd(\Ds-1)\p_{a}\psi \p^{a}\te\right]
\ee
and
\be
V_{\mt{grav}}=\4\left[ 2\Dz e^{-2\psi}-\2\Dz e^{2(\te-\psi)}+k\Dd e^{-2(\psi+\te)}\right]
\ee
One can also rewrite the cosmological and the flux terms.
\be
S_{\mt{cos}}=-2\hat{\La} \frac{M_p^2}{2}\int dx^{D+1}\ e^{-1}e^{\Ds\psi+\Dd\te}
\ee
and
\be
S_{\mt{flux}}=-\4 M_p^2\int dx^{D+1}\ e^{-1}\left[\cs^2e^{-\Ds\psi-\Dd\te}+\cd^2e^{(\Ds-2\Dd)\psi-\Dd\te}\right]
\label{eq:4c-flux}
\ee

We now perform the conformal rescalings:
\be
e_a{}^m=\Delta e'_a{}^m\ ;\ \Psi=(\Delta)^{-1}\Psi'\ ;\ \Delta=\Psi'^{\frac{\Ds}{\Dh-2}} \Te'^{\frac{\Dd}{\Dh-2}} \,,
\ee
The total effective action then reads
\be
S_{\mt{eff}}=S_{\mt{grav}}+S_{\mt{cos}}+S_{\mt{flux}}=\frac{1}{2}\int d^{D+1}x\ e^{-1}[R+K-V],
\ee
where
\be
K=-\frac{1}{\Dh-2}\biggl[\Ds(D-1)(\p\psi)^2+\Dd(\Dz+D-1)(\p\te)^2+2\Dd(D-1)\p_{a}\psi \p^{a}\te\,\biggr]
\ee
$$
V_{\mt{eff}}=V_{\mt{grav}}+V_{\mt{cos}}+V_{\mt{flux}}=-\frac{1}{4}\biggl[2\Dz e^{-2\psi}-\frac{1}{2}\Dz e^{-2(\psi-\te)}+k\Dd e^{-2(\psi+\te)}\biggr]$$
\be+2\widehat{\La} e^{-2\al(\psi+\nu\te)}+\2\left[\cs^2e^{-2\beta(\psi+\nu\te)}+\cd^2e^{-2(\ga\psi+\de\te)}\right]\,.
\ee
with
$$\nu=\frac{\Dd}{\Ds}$$
$$\al=\frac{\Ds}{\Dh-2}$$
$$\beta=\frac{\Ds.D}{\Dh-2}$$
$$\ga=\frac{\Dd D+\Dz}{\Dh-2}$$
and
\be
\de=\frac{\Dd(D+\Dz)}{\Dh-2}
\ee
{\bf Diagonalization and Normalization:} In the earlier subsection we obtained the complete effective action for the moduli scalars coupled to gravity. For convenience in the cosmological analysis, it is useful to diagonalize the kinetic terms in the action. Such a diagonalization can  be acheived by choosing the volume ($v$) and the shape ($\theta$) as variables
$$v=\psi+\nu\te$$
After diagonalization we have
\be
K=-\frac{1}{\Dh-2}\left[\Ds(D-1)(\p v)^2+\frac{\Dd\Dz}{\Ds}(\Ds+D-1)(\p \te)^2
\right]
\ee
while the potential in terms of this new variable looks like
\be
V=\4e^{-2v}\left[\frac{\Dz}{2}e^{2(\nu+1)\te}-\Dd.ke^{2(\nu-1)\te}-2\Dz e^{2\nu\te}\right]
+2\hat{\La}e^{-2\al v}+\frac{\cs^2}{2}e^{-2\beta v}+\2\cd^2e^{-2(\ga v+\de'\te)}
\label{3potential}
\ee
\be
\equiv \sum_{i=1}^6V_{i}e^{X_i}
\ee
where
\be
\de'=\de-\ga\nu=\frac{\Dz\Dd}{\Ds}
\ee
For cosmological analysis it is also convenient to rescale  the fields
\be
v\ra K_v v \mx{ and } \te\ra K_{\te}\te
\ee
with
\be
K_v=\sqrt{\frac{\Ds(D-1)}{\Dh-2}}\mx{ and }K_{\te}=\sqrt{\frac{\Dz\Dd}{\Ds}}
\label{3scale}
\ee
so that the coefficient in front of the kinetic terms is normalized to 
 the canonical value of 1/2. This essentially has the effect of rescaling the exponents for the fields $v$ and $\te$ by $K_v^{-1}$ and $K_{\te}^{-1}$ respectively. 
\vs
{\bf Couplings:} We have successfully obtained the classical effective potential (scalar interactions) for our moduli fields. However they will in general also couple to matter and radiation. These couplings will depend on whether the matter-radiation fields are residing on the brane~\cite{Sundrum} or on the bulk~\cite{kaluza}. We focus on the radiation\footnote{Obtaining fermionic couplings are more complicated as the fermionic representations can be different depending upon the number of extra dimensions. Also, four dimensional fermions can come from higher dimensional gravitinos. We leave this topic for further study.}  term and obtain the couplings for the two different cases: \\
1. Bulk Radiation: Fluctuations of the off-diagonal part of the full higher dimensional metric can be identified as four dimensional gauge fields. These fields will in general couple to the moduli fields and the coupling can be obtained following the usual dimensional reduction procedure \cite{kaluza}. One finds that after the conformal rescalings the action for the gauge fields to be given by
\be
S_{bulk,gauge}=-\8 \int\ d^{D+1}x\ e^{-1}e^{2(\psi+\te)}F^2=-\8 \int\ d^{D+1}x e^{-1}e^{2(v+(1-\nu)\te)}F^2
\label{3bulk}
\ee
where $F_{mn}^{\ad}$ is the usual field strength for the unbroken (non-)abelian gauge fields $A_m^{\ad}$. \\
2. Brane Radiation: In the brane-world picture the action for radiation localized on the brane is given by
\be
S_{brane,gauge}=-\4 T\int\ d^{4}x\ \sqrt{-\gh(\mt{brane})}F_{mn}F_{pq}\gh^{mp}\gh^{nq}
\ee
where $T$ is the brane tension in Planckian units.
Now $\gh(\mt{brane})$ is related to $g(\mt{brane})$ through conformal rescalings
$$
\sqrt{-\gh(\mt{brane})}=\sqrt{-g(\mt{brane})}\Delta^{-(D+1)}$$
while
$$\gh^{mp}=\Delta^2g^{mp}$$
and thus the $\Delta$ dependence completely cancels in the brane action for $D=4$, which is only a consequence of the conformal invariance of Yang-Mill's action in four dimensions.
Then
\be
S_{brane,gauge}=-\4 T\int\ d^{D+1}x e^{-1}F^2
\label{3brane}
\ee
i.e. there is no direct coupling of the moduli fields to brane radiation term.
\setcounter{equation}{0}
\section{{\bf COSMOLOGICAL EVOLUTION}}
{\bf Exponential Potentials and the Mapping Technique:} Having obtained the effective potential let us try and analyze what it looks like in the $v-\te$ plane. To have a qualitative understanding of the potential we develop a ``mapping technique'' which works in general for a potential with sums of exponentials of two fields. We start by looking  at how the potential behaves as one goes asymptotically to infinity in any direction, say
$$\te=mv$$
Since the potential is  a sum of  exponentials it is clear that as one proceeds to infinity along a particular direction only one of them, the one with the largest (or smallest) ``effective exponent'' will dominate as $v\ra +\infty$ (or $-\infty$). Thus in the $v-\te$ plane one can map out regions where the different exponents dominate. Moreover, depending upon the sign of the dominant effective exponent and the coefficient in front of the dominating exponential potential, the potential will either increase (``wall''), decrease (``fall'') or asymptotically approach zero (``plane''). The map therefore provides a  qualitative picture of the dynamics. Since there are six exponentials in our model one has to compare six different effective exponents for a given slope, as $v\ra \pm\infty$:
$$
X_1=-2+2(1+\nu)m$$
$$X_2=-2+2\nu m$$
$$X_3=-2-2(1-\nu) m$$
$$X_4=-2\al$$
$$X_5=-2\beta$$
\be
X_6=-2(\ga+m\frac{\Dz\Dd}{\Ds})
\label{4exponents}
\ee
Also note
\be
\beta>\ga>\al\mx{ while } 1>\nu>0
\ee
Comparing the different exponents in (\ref{4exponents}) one can obtain ranges of $m$ where a particular exponent dominates. For example, for $m>0$ and $v\ra \infty$ it is clear that $X_1>X_2>X_3$. While 
$$X_4>X_1\Ra -2\al>-2+2(1+\nu)m\Ra m<\frac{1-\al}{1+\nu}$$
i.e. $X_4$ overtakes $X_1$ when $m$ becomes less than $(1-\al)/(1+\nu)$. Proceeding in this manner one can indeed map out the entire $v-\te$ plane.
Computing the sign of the dominant  exponent also tells us  how the potential looks as one goes towards infinity. Consider the $X_1$ exponent:
$$X_1>0\Ra -2+2(1+\nu)m>0\Ra m>\frac{1}{1+\nu}$$
Thus when $m>1/(1+\nu)$ the potential increases exponentially  (the coefficient in front being positive) so that we hit a wall. However, when $m<1/(1+\nu)$ then the exponent is negative, so that the potential asymtotically falls off to zero and the same thing happens as $X_1$ gives way to $X_4$. Proceeding in this manner once we know the entire $v-\te$ map, it is easy to understand the dynamics, specially  when one is sufficiently away from the origin\footnote{The origin gets its special status because we assume that all the mass parameters are of order 1, i.e all the potentials are roughly of equal strength near the origin. If this is not the case, the origin essentially gets shifted, but the analysis still holds.}. Then essentially the ``particle'' rolls along a single (the dominant) exponential potential and such behaviour  has been extensively studied \cite{liddle} in the literature. 

\begin{figure}[!h]
\begin{center}
\includegraphics[scale=0.9,angle=0]{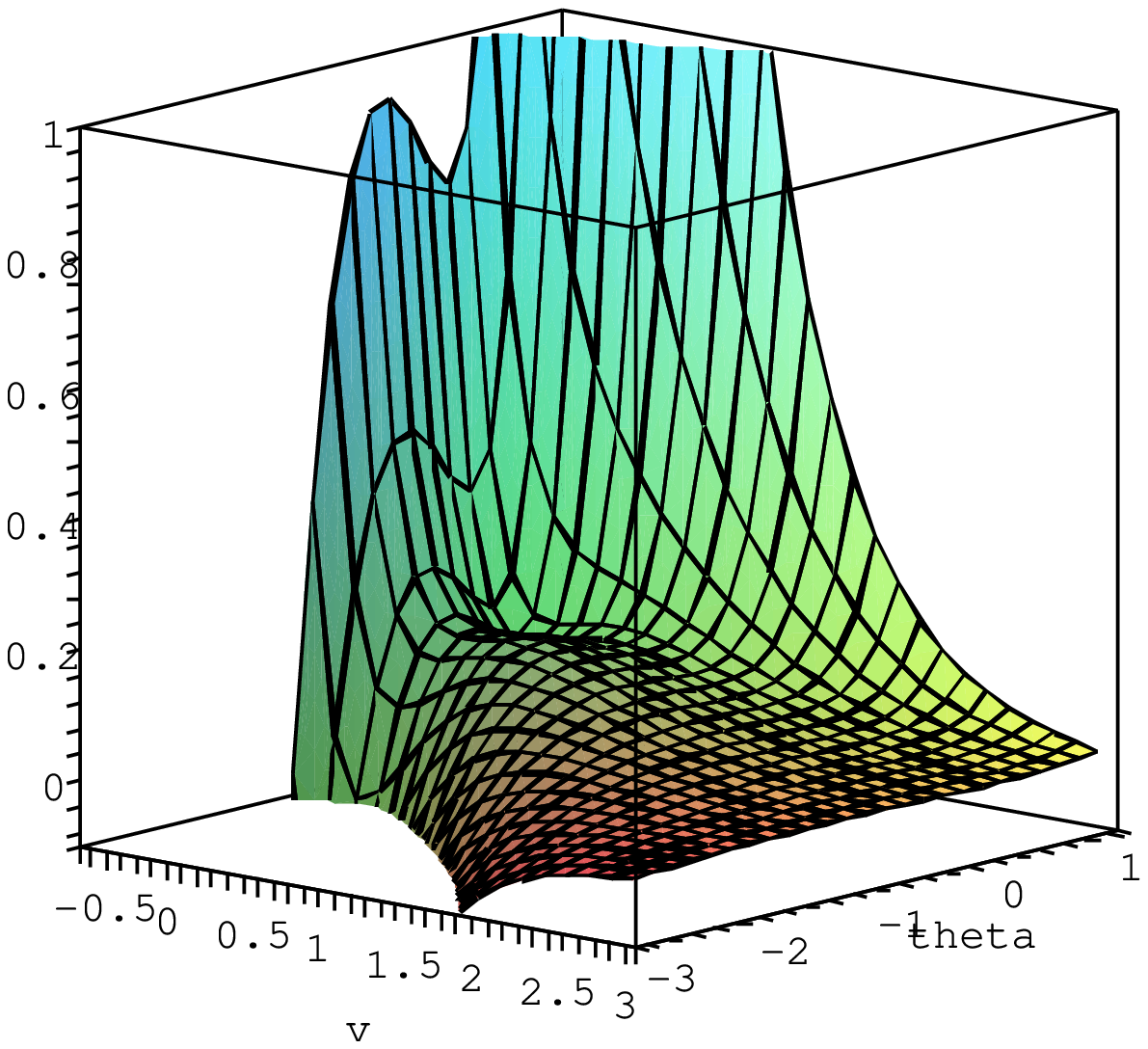}
\end{center}
\caption{\small The potential with $\cd~=~0$.}
\label{figps1}
\end{figure}

If the dominating exponential term is given by
\be
V_{\mt{eff}}\approx Ve^{\al v+\bb\te}
\ee
with $V$ being the coefficient in front then the particle moves along a constant direction given by 
\be
\vec{u}=-\frac{V}{|V|}(\al\hat{v}+\bb\hat{\te})
\label{4velocity}
\ee
under the influence of an effective exponent
\be
\al_{\mt{eff}}=\sqrt{\al^2+\bb^2}
\ee
Now it is known \cite{liddle} that $\al_{\mt{eff}}<\sqrt{2}$ indicates a scalar field dominated regime of accelerated expansion while $\al_{\mt{eff}}>\sqrt{2}$ corresponds to matter-radiation domination where the scalar field tracks matter-radiation densities. So let us find out whether the exponents we have support acceleration or not. To obtain bounds on the effective exponents, we of course  rescale the exponents  using  (\ref{3scale}). Playing around with the dependence of the various  dimensions in the exponents we find\footnote{There may be stronger bounds but these are sufficient for our purpose.}  
$$X_{1\mt{eff}}>\sqrt{2}$$
$$X_{2\mt{eff}}>\sqrt{2}$$
$$X_{3\mt{eff}}>\sqrt{2}$$
$$\sqrt{2}>X_{4\mt{eff}}\geq \sqrt{\frac{6}{5}}$$
$$3\sqrt{2}>X_{4\mt{eff}}\geq 3\sqrt{\frac{6}{5}}>\sqrt{2}$$
\be
X_{6\mt{eff}}>2
\ee
It is clear that the only region which generically supports acceleration is when $X_4$ dominates. This is not surprising given its origin in the higher dimensional cosmological constant but what is important in our model is that, we do not need to start with a  ``small'' cosmological constant. One should be cautioned that this behaviour is only expected at late times if and when the particle is well ``inside'' the region when one can approximate the potential by a single exponent. The particle can have interesting transient behaviour \cite{transient} when it is moving from a region dominated by one exponent to a  region dominated by a different exponent. In our model, one example would be when the particle enters $X_4$-region from $X_1$-region. Under certain conditions the particle may also  follow a different attractor solution which is a compromise between the two attractor solutions corresponding to the two competing adjacent exponents. Such situations will arise in our model with interesting consequences and we will elaborate on them later. To begin with let us try and understand the generic picture with the help of the $v-\te$ map. 
\vspace{5mm}
\\ 
{\bf Quintessence or Moduli Stabilization:} To understand the effect of the fluxes we first look at the potential when we have not turned on the flux along the subgroup. In this case (see fig.~\ref{figps1}) the minimum in the potential corresponds to the symmetric internal manifold (gauge symmetry unbroken) which  makes it interesting for studying possible symmetry breaking mechanisms \cite{t2,t1}.

\vskip 0.1cm

We note that the moduli in the lower half plane is not stabilized and in particular the potential falls to $-\infty$ in the region bounded by the straight lines 
$$\te=\frac{\bb-1}{1-\nu}v\mx{ and }\te=-\frac{1}{1-\nu}v$$ 
essentially because $X_3$ dominates in this region.
Thus to perform meaningful cosmology one has to be able to avoid the ``fall'' which may be possible depending upon the initial conditions. Before we investigate this any further though, let us see what happens if we now turn on the flux along the subgroup. \\
\begin{figure}[!h]
\begin{center}
\includegraphics[scale=0.9,angle=0]{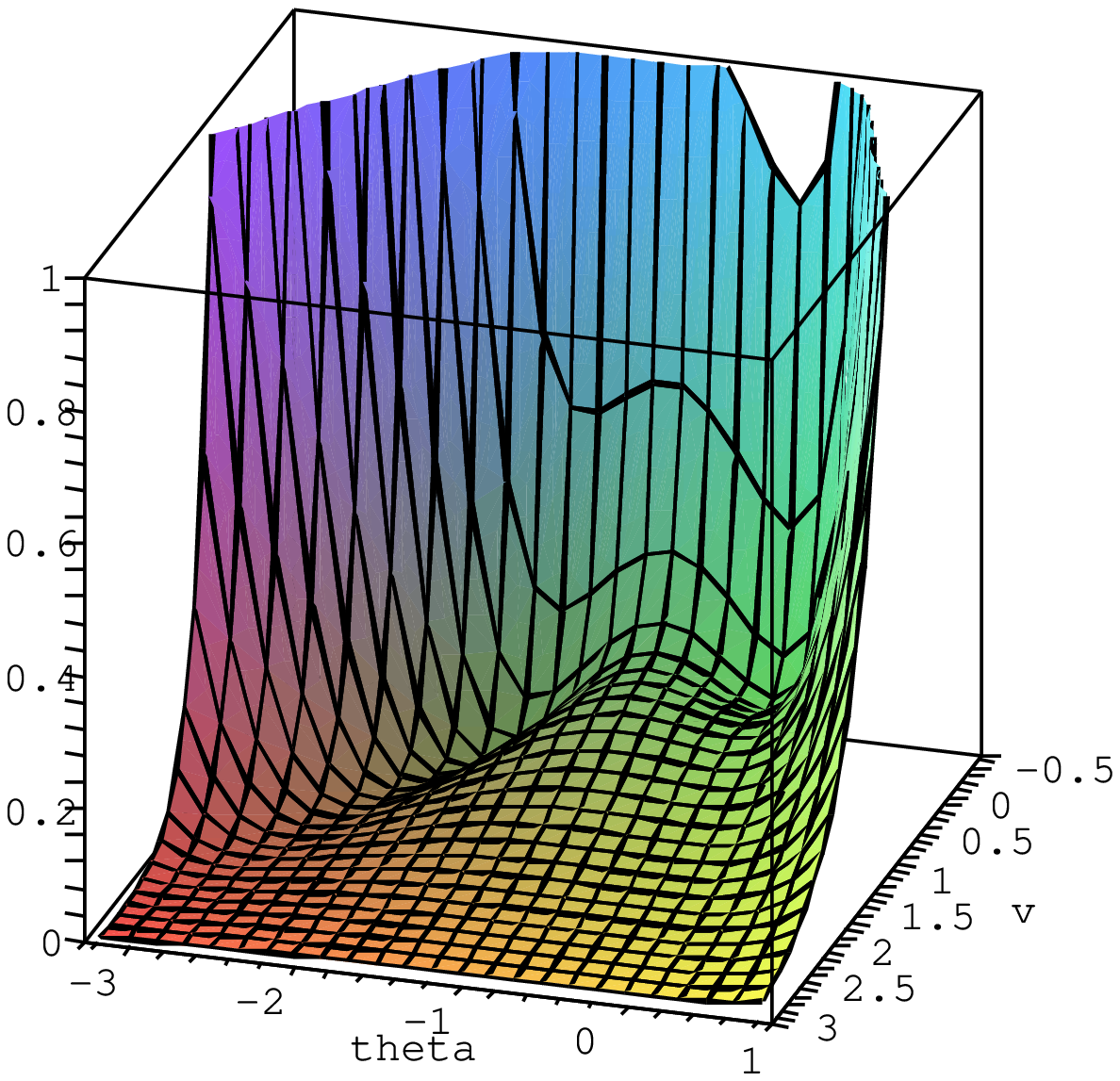}
\end{center}
\caption{\small The potential with $\cd~\neq~0$.}
\label{figps2}
\end{figure}
\vskip 0.1cm
The potential looks  (see fig.~\ref{figps2})  similar to the previous case except that we find the shape moduli field $\te$ now has a stable potential, perhaps not surprisingly. Previously  it has been argued in the context of string theory and Calabi-Yau manifolds how flux should stabilize the shape moduli fields \cite{kachru}. Our model furnishes an explicit realization of the same, albeit in the context of an internal group manifold. 

\begin{figure}[!h]
\begin{center}
\includegraphics[scale=0.45,angle=0]{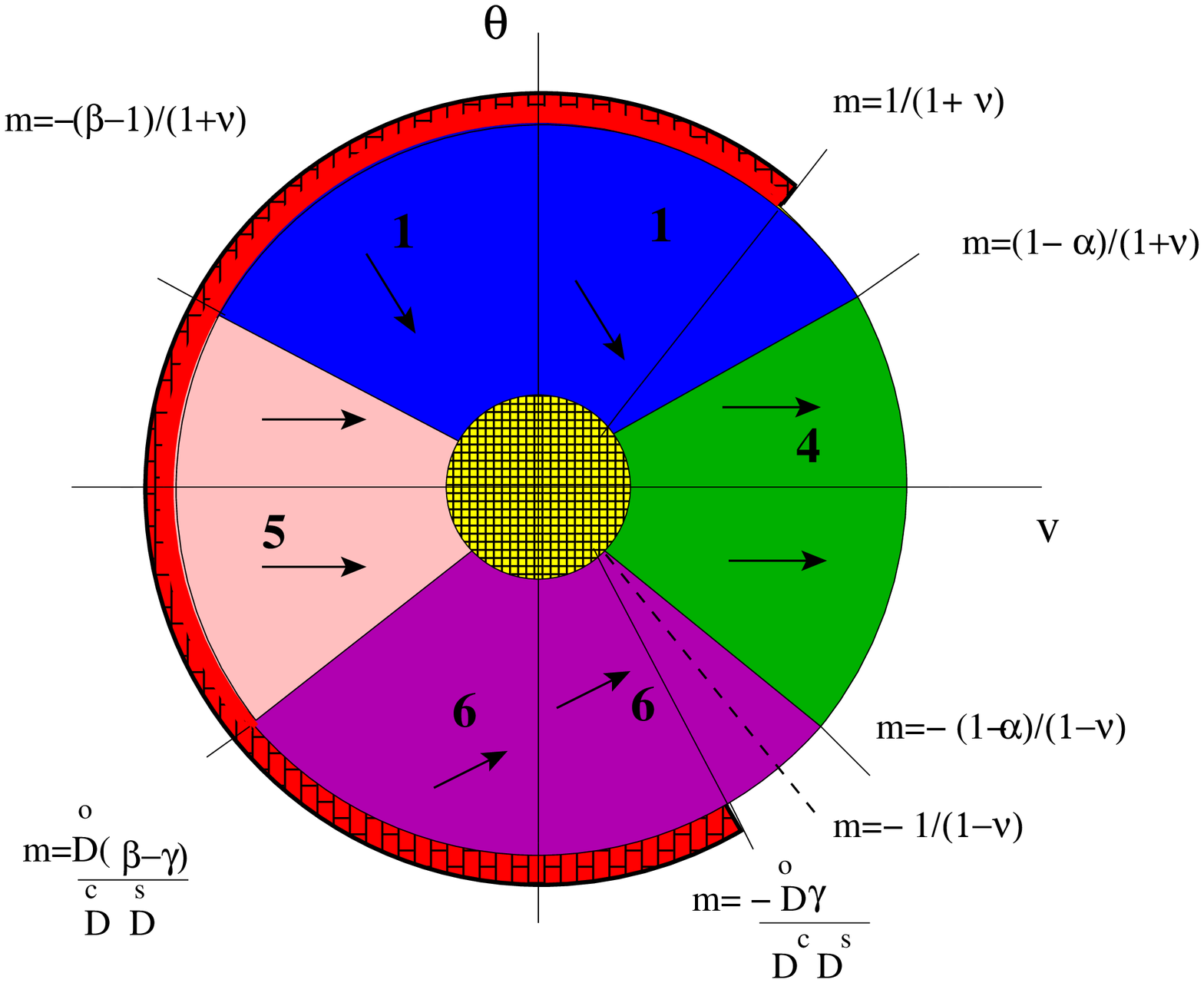}
\end{center}
\caption{\small The $v-\theta$ map for $k>0$. The solid radial lines demarcate regions
where different exponentials dominate, the arrows indicating the direction in which the particle moves. The dashed line corresponds to $\alpha_{\mt{fine}}=$~constant.
}%
\label{figps3}
\end{figure} 
\vskip 0.1in

Let us now focus on the cosmological evolution of the moduli fields. We will consider the most general case when all the coefficients of the exponents are non-zero. The evolution when some of the coefficients are zero (like when the subgroup flux is not turned on) can mostly be derived from the general case, although  interesting differences sometimes emerge and we will comment on them as we go along.   One can get a good idea of the evolution by studying the $v-\te$ map (see fig.~\ref{figps3}) but the evolution clearly depends on the initial conditions.  The {\bf I}nitial conditions can be broadly classified under two categories: either (I-1)  our universe starts from the symmetric minimum (see figure) and then may evolve to a broken phase, or (I-2) it starts out far away from the minimum in any particular direction. Among I-2 cases for realistic cosmological scenarios, we consider the particle to be initially rolling  down a wall ($X_6,X_5$ or $X_1$). Let us consider these situations in a little more detail.\\
I-1. There are  two possibilities depending on  how deep the hole (potential minimum) is.\\
(a) The tunnelling probability may be so small that the particle remains trapped for the entire evolution of our universe thus far; there will be no interesting cosmology to be gained but at least we would have solved the moduli stabilization problem\footnote{Observe that the internal manifold will be trapped in its symmetric phase which often preserves some amount of supersymmetry \cite{kaluza} and hence has a chance to address the cosmological constant problem.}.  Note a subtle but interesting difference when the extra flux is turned on: the minimum is no longer at the symmetric $\te=0$ point. It is slightly shifted indicating a small squashing. In other words if the hole is deep then the universe will be stuck in a slightly squashed or broken phase. As observed in \cite{t1,t2} the mass of the broken gauge bosons depend very sharply on the squashing and in order to produce a large hierarchy the squashing has to be very small, which may  be achieved in this scenario. \\
(b) If the hole is shallow then the particle may be able to get out of it  by tunneling. It is however well known \cite{linde} that the collision of bubbles formed as the universe undergoes a phase transition through tunnelling produces inhomogeneities that are too large as compared to observation and hence we do not consider this possibility in any detail here. However, if the hole is really shallow  so that a small velocity-kick originating from classical or quantum fluctuations  may induce the phase transition then a realistic inflationary scenario may emerge as was considered in \cite{t2}. We should point out that when one introduces the additional flux, then for certain parameter ranges, the potential can develop a second minimum. If this second minimum is stable then the inflaton can oscillate around it to reheat the universe \cite{linde} and eventually settle down offering us a scenario where at {\bf L}ate times all the moduli are stabilized (L-3). Alternatively, one can imagine a hybrid type inflationary model~\cite{hybrid} where as the inflaton approaches the minimum, it veers off in another ``steeper'' direction (and thereby ending inflation) and ultimately to either L-1 or L-2. This is reminiscent of the quintessential inflation models \cite{peebles} where nevertheless the role of inflaton and quintessence field  is played by different fields. We will study these in some detail in the next section.\\
I-2. From the $v-\te$ map and the arrows  in the map (see fig.~\ref{figps3}) one sees that, in general, along the upper half plane one moves from  the $X_5$-region  onto the  $X_1$-region (or starts from $X_1$ region) and then finally to the $X_4$-region. Note, since neither $X_5$ nor $X_1$ supports acceleration, the universe while passing through these regions would evolve as in the radiation-matter dominated phase. This will eventually be followed by an accelerated phase in the $X_4$-region which can continue for eternity and we will call this as the L-1 {\bf L}ate time behaviour. Because of the presence of the extra flux, the evolution in the lower half is no longer pathological. One typically moves from the $X_5$-region to the $X_6$-region (or starts from $X_6$ itself) and then finally  to the $X_4$-region where we can have quintessence. Again, since $X_6$ also does not support acceleration, the generic scenario is radiation-matter domination followed by quintessence. 

Thus, for I-2 we seem to have a history of our universe consistent with observation; radiation-matter domination followed by a phase of quintessential acceleration which presumably we are observing today. It should be mentioned that in \cite{ohta} it was also observed that one could have late time  acceleration phases  for compact  internal manifold when relevant fluxes are turned on. However in order to have a viable quintessence scenario one has to also satisfy the cosmological observational bounds on the quintessence equation of state 
and the cosmological constant energy density.
Further, if the quintessence field couples to matter/radiation, observational bounds coming from time variation of various coupling constants and fifth force experiments \cite{review} need to be satisfied and we will investigate all these in the next section. 

With the help of the maps we have so far discussed qualitatively some likely evolution of the universe. One thing we should mention  is that it is not necessary to assume that our current universe lies in the  $X_4$-region, although this is the only region which supports acceleration. The reason is transient acceleration may potentially  be able to explain the observed phenomenon of quintessence \cite{transient}. For example, we can be caught up in the transition between the $X_5$-region and the $X_6$-region, or between say, the $X_5$-region and the $X_1$-region. One can also have  other viable late time accelerating phases as we will see in the next subsection.
\vs\\
{\bf Dynamics with Two Exponential Potentials:} We have seen that for a wide range of initial conditions SUGRA type models give rise to a radiation-matter dominated phase followed by quintessence. However one can also obtain an early phase of acceleration (inflation) from the dynamics near the well. As mentioned earlier, if our universe starts from the symmetric minimum and   the potential well is shallow  then the universe could inflate \cite{t2} while getting out of the well via an initial velocity kick. Here we will find that it is not necessary for the universe to start from the well but that there are mechanisms available which can lead the particle towards the well even if one starts with initial condition I-2.  This can happen if the particle is trapped in a ``trough'' created by two competing exponents as we will illustrate. Thus, the I-2-type initial conditions further divides into two categories: either the particle follows the original trajectory described earlier and we will continue to refer such initial conditions by I-2, or the particle can be attracted towards the well, type I-3 initial conditions. In this case  its further dynamics proceeds in a manner similar to type I-1 initial conditions, so in some sense I-3 is  a mix of I-1 and I-2. To understand how this happens  we need to analyze the dynamics of the particle when there are two adjacent competing exponential potentials. As a bonus we will also find a second late time accelerating attractor solution!
\begin{figure}[!h]
\begin{center}
\includegraphics[scale=0.9,angle=0]{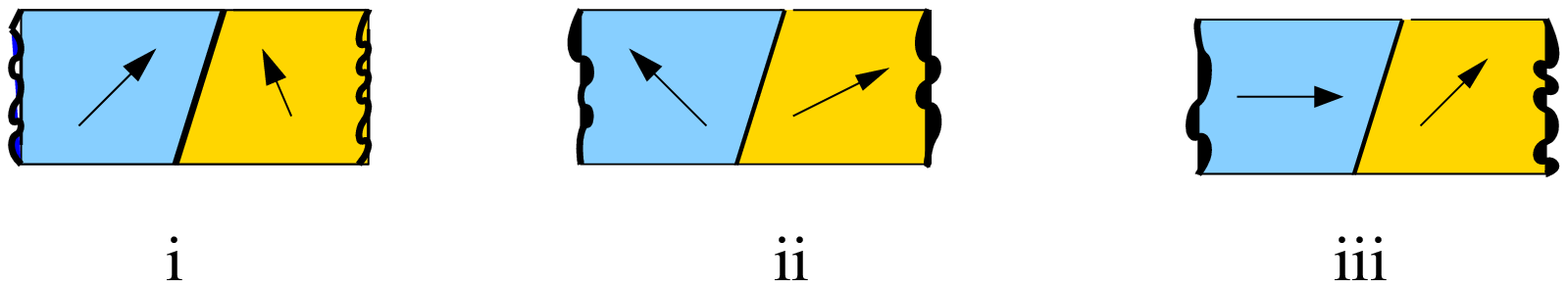}
\end{center}
\caption{\small Three different late-time behaviours depending on the orientations in competing regions (see text for details).}
\label{figps4}
\end{figure} 
\vskip 0.1in
 For simplicity, let us consider a potential containing two exponentials:
\be
V=V_1e^{\al_1v+\bb_1\te}+V_2e^{\al_2v+\bb_2\te}\equiv \sum_{i=1}^2V_ie^{X_i}\,;\quad \bb_1>\bb_2
\ee
As our earlier analysis showed, the $v-\te$ map gets divided into two regions, where the two different exponents dominate (see fig.\ref{figps4}). Now, depending  upon the direction of motion in these two regions which can be computed from (\ref{4velocity}), three different kinds of dynamics or attractor behaviour is possible as illustrated by fig.~\ref{figps4}. In  case (i) the arrows point towards each other so that a ``trough'' is created between the two regions. So no matter where the particle initially starts from it will ultimately end up in the trough and continue to move exactly along the boundary demarkating the two competing exponential regions \cite{liddle}: 
\be
\vec{u}_{\mt{comb}}=\pm(\bb_2-\bb_1)\hat{v}\mp(\al_2-\al_1)\hat{\te}
\label{4combv}
\ee
The particle evolves as if governed by an effective exponent given by
\be
X_{\mt{eff}}=\left|\frac{\al_1\bb_2-\al_2\bb_1}{\sqrt{(\al_2-\al_1)^2+(\bb_2-\bb_1)^2}}\right|
\label{4comb}
\ee
Thus it may happen that even though the individual exponents are too large to support inflation the combined exponent (\ref{4comb}) may, and this was already noted in some special cases in \cite{anupam}. 

\vskip 0.1cm

Coming back to the different cases, in (ii) we see that the arrows are pointing in opposite directions. Such a situation implies an unstable ridge between the two regions. So there are two late time attractor solutions. The particle either slides to the $X_1$-region or $X_2$-region depending on the initial conditions and evolves as if the other exponent ($X_2$ and $X_1$ respectively) is absent. 

In (iii) one of the arrows points towards the other region but the second arrow points towards its own region. The dynamics is again simple. The second arrow is the late time attractor. The particle may evolve in the first region for a while according to the exponent $X_1$ but eventually it encounters $X_2$-region, after which it makes a transition to the $X_2$ late time attractor solution. The dynamics during such a transition may also be quite interesting as pointed out in \cite{transient}. 
\vs
{\bf Phases of Acceleration :} Let us investigate when in our model one can obtain  troughs of type (i). In order to do this first we need to figure out the directions ($\vec{u}_i$) of the trajectories in the different regions and then check when we get type (i) for the different adjacent regions that are relevant. Here is the list:
$$\vec{u}_1=2(\hat{v}-(1+\nu)\hat{\te})$$
$$\vec{u}_2=-2(\hat{v}-\nu\hat{\te})$$
$$\vec{u}_3=-2(\hat{v}+(1-\nu)\hat{\te})$$
$$\vec{u}_4=2\al\hat{v}$$
$$\vec{u}_5=2\bb\hat{v}$$
\be
\vec{u}_6=2(\ga\hat{v}+\de'\hat{\te})
\ee
Clearly, the ``velocities''  $\vec{u}_1$ and $\vec{u}_6$ imply a transition of  type (iii) between $X_1-X_4$ and $X_6-X_4$ (see fig.~\ref{figps3}), and thus the late time behaviour is dominated by $X_4$. Let us now compare the  velocities $\vec{u}_1$ and $\vec{u}_6$ with the slopes demarcating  the $X_5-X_1$ and $X_5-X_6$ region respectively. One finds that we have type (iii) situation if the various dimensions satisfy
\be
\frac{(\Ds-\Dd)\Dd^2(\Ds+2)^2}{(\Ds+\2\Dd)\Ds^2}<2
\ee
and
\be
\frac{(\Ds+\Dd)^2(\Ds+2)}{(\Ds-1)\Ds^2}<2
\ee
respectively and the dynamics proceed as in I-2. Otherwise, depending upon the initial conditions as the particle nears  the $X_5-X_1$ or $X_5-X_6$ boundary it will be sucked into the $X_5-X_1$\footnote{We note in passing that for large number of extra dimensions, $\Ds>30$, the $X_5-X_1$ track supports acceleration and can therefore account for a phase of inflation.} or $X_5-X_6$ combined track (type i) and  the particle will enter the ``near-well'' region; we refer to this case as I-3.
 
\vskip 0.1cm

We further observe that  near the well, the exponents $X_2$ and $X_3$ kick in, which  can be instrumental in creating an attractor behaviour in conjunction with other exponents. Suppose  $k>0$. Then as one approaches inward towards the well in the  $X_6$-region, the $X_3$ exponent become strong as can be seen from (\ref{4exponents}). A trough or a valley  is created, sandwiched between the $X_6$ and so to speak, the $X_3$-region, as is also evident in the picture (fig.~\ref{figps3}) of the potential. One can compute the effective exponent ($X_{36}$) in this valley.  
\be
X_{36}=\sqrt{\frac{2\Dz}{\Dz+2}}<\sqrt{2}
\ee
Thus this can indeed support acceleration. This phase of acceleration gives us an alternative late time  quintessence phase, L-2. We observe that if the higher dimensional cosmological constant is absent, then this trough will actually be the generic late time attractor whereas if $k=0$ then L-2 is absent.

To summarize, we have three initial conditions I-1,2,3 and two late time quintessence evolution, L-1,2, plus a stable L-3 late time configuration. For I-2 $\ra$ L-1,2 the dynamics is clear; radiation domination followed by quintessence. However for I-1,3 $\ra$ L-1,2,3 there is an intermediate evolution in the vicinity of the well about which we about which we try to shed some light in the next section. 
\section{{\bf REALISTIC SCENARIOS AND CONSTRAINTS}} 
{\bf Interesting Cases:} We now have a qualitative understanding of the different trajectories that the particle may follow, as well as its cosmological significance, but the possibilities are numerous. So here let us focus on the likely and interesting ones. First of all, let us assume that the well is shallow enough so that the particle can roll out of it. Second, we realize that depending on the  initial conditions and the various parameters of the potential, the particle/inflaton, after crossing the  potential barrier, can enter a region dominated either by $X_1$, $X_3/X_6$ or $X_4$. In the first two cases inflation can end as neither exponents support acceleration; however, it doesn't if the inflaton directly goes over to the $X_4$-region. Hence the realistic scenarios can be summarized as: 
\begin{center}
\vspace{5mm}
\begin{tabular}{||l|l|l|l||}\hline
{\bf Case}& {\bf I.C.}& {\bf Evolution}& {\bf Phases}\\ \hline
1. & I-1  & W$\ra  X_1\ra X_4$      & E$\ra$R$\ra$L-1\\ \hline 
2a.& I-1  & W$\ra W_2$    & E$\ra$L-3\\
2b.& I-1  & W$\ra X_3\ra X_{36}$    & E$\ra$R$\ra$L-2\\
2c.& I-1  & W$\ra X_3,X_6\ra X_4$   & E$\ra$R$\ra$L-1\\ \hline
3a.& I-2  & $X_5\ra X_{1}\ra X_4$   & R$\ra$R$\ra$L-1\\
3b.& I-2  & $X_5\ra X_{6}\ra X_4$   & R$\ra$R$\ra$L-1\\ 
3c.& I-2  & $X_5\ra X_{6}\ra X_{36}$& R$\ra$R$\ra$L-2\\ \hline
4a.& I-3  & $X_5\ra X_{15}\ra$1. or 2.& R$\ra$R$\ra$1. or 2.\\
4b.& I-3  & $X_5\ra X_{65}\ra$1. or 2.& R$\ra$R$\ra$1. or 2.\\ \hline
\end{tabular}

\vspace{5mm}
\end{center}
where $W_2$ labels a stable second minimum, E denotes a possible early phase of inflation and R corresponds to radiation-matter domination. As is evident from the table, most  of these scenarios possess a late time acceleration phase which may be able to explain the observations today, while some of them admit an early phase of acceleration or inflation. In the next subsections, we investigate in details whether these phases of acceleration can indeed be compatible with the various observational bounds coming from early and late time cosmology as well as particle physics.   
\vs
{\bf Constraints from Inflation:}  For an inflationary scenario to be  successful it should be able to produce at least 50-60 efoldings to explain the standard cosmological issues presented by the flatness and horizon problems \cite{linde}. The inflation scale should be such that it generates CMB fluctuations with the right amplitude ($\delta_{\mt{H}}$) \cite{linde}
\be
\delta_{\mt{H}}\sim10^{-5}
\label{5cmb}
\ee

\begin{figure}[!h]
\begin{center}
\includegraphics[scale=0.45,angle=0]{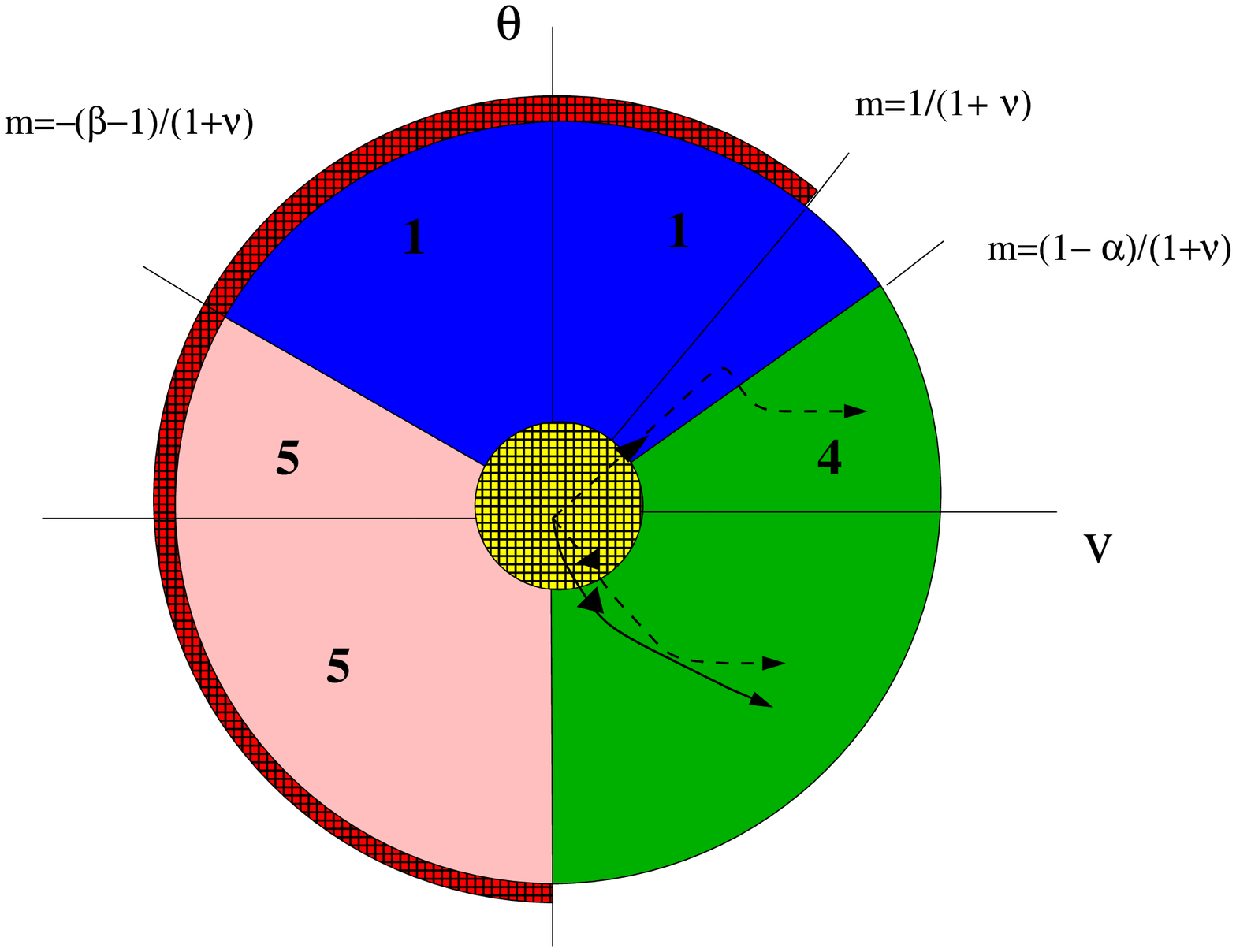}
\end{center}
\caption{\small The $v-\theta$ map for $k=0$. The dashed lines correspond to straight line approximations to the more realistic (solid line) evolution.}
\label{figps5}
\end{figure}

Now, to answer these questions comprehensively one has to perform a detailed numerical analysis involving evolution of both fields, but one can get significant insight into the physics by employing approximate analytical methods as we will show. We start by ``freezing'' a linear combination of the fields \cite{t2}, i.e.  assume that the inflaton moves along a constant slope in the $v-\te$ plane:
\be
\sigma=v+m^{-1} \te=\mt{ constant} 
\label{eq:stabilization}
\ee 
and for analytic simplicity we choose 
\be
m_{\mt{infl}}^{-1}\approx 0
\ee
 The subsequent evolution of the inflaton can also be approximated by straight line trajectories the slope of which is determined by the specific regions it is traversing (refer to table). A simplified picture of evolution would thus look like fig.~\ref{figps5}. Of course in reality, these straight line evolutions will be joined by smooth curves.

\vskip 0.1in

We start our analysis with the potential (\ref{3potential}) which can be rewritten as
$$
V=\4e^{-2v}\left[\frac{\Dz}{2}e^{2(\nu+1)\psi}-\Dd.ke^{2(\nu-1)\psi}-2\Dz e^{2\nu\psi}+\Ds\right]$$
$$+2\hat{\La}e^{-2\al v}-\4\Ds e^{-2v}+\frac{\cs^2}{2}e^{-2\beta v}$$
$$\equiv \4e^{-2v}\ti{V}(\te)+V_{\mt{min}}(v)$$
where we have also made the approximation that near the well, the $X_6$ contribution is negligible. It is known that the minimum of the potential $\ti{V}(\te)$ occurs at 
$$\te_{\mt{min}}=0\mx{ and } \ti{V}(\te=0)=0$$
so that the total potential is minimized by minimizing $V_{\mt{min}}(v)$. One can compute this approximately. In general for a potential of the form
\be
V \approx V_1 e^{\al_1 v}-V_2e^{\al_2 v}+V_3 e^{\al_3 v}
\ee
with
\be
\al_1>\al_2>\al_3
\ee
$$
V'=V_1\al_1 e^{\al_1 \phi}-V_2\al_2e^{\al_2 \phi}+V_3\al_3 e^{\al_3 \phi}=0$$
$$\Ra e^v\approx\left(\frac{V_2\al_2}{V_1\al_1}\right)^{\frac{1}{\al_1-\al_2}}$$
and
\be
 V_{\mt{min}}\approx V_2 \left(\frac{\al_2}{\al_3}-1\right)\left(\frac{V_2\al_2}{V_1\al_1}\right)^{\frac{\al_2}{\al_1-\al_2}}
 \label{5vmin}
 \ee
With some algebraic manipulations we find in our case
\be
V_{\mt{min}}=\2 e^{-2v_{\mt{min}}}= \2\left(\frac{\Ds+2}{12V_{\mt{flux}}}\right)^{\frac{\Ds+2}{2(\Ds-1)}}
\ee
Thus if $V_{\mt{flux}}$ is large compared to typical $V_{\mt{curv}}$, the typical hierarchy required being $10^4$, then $V_{\mt{min}}$ is much smaller as compared to $M_p$ and produces the right size of CMB fluctuations (\ref{5cmb}). For example say $\Ds=2$ and $V_{\mt{flux}}\sim 10^4$. Then
$$\delta_{\mt{H}}\sim \sqrt{V_{\mt{min}}}\sim\sqrt{\left(\frac{4}{12\times10^4}\right)^{2}}\sim 10^{-5}$$

\begin{figure}[!h]
\begin{center}
\includegraphics[scale=0.4,angle=0]{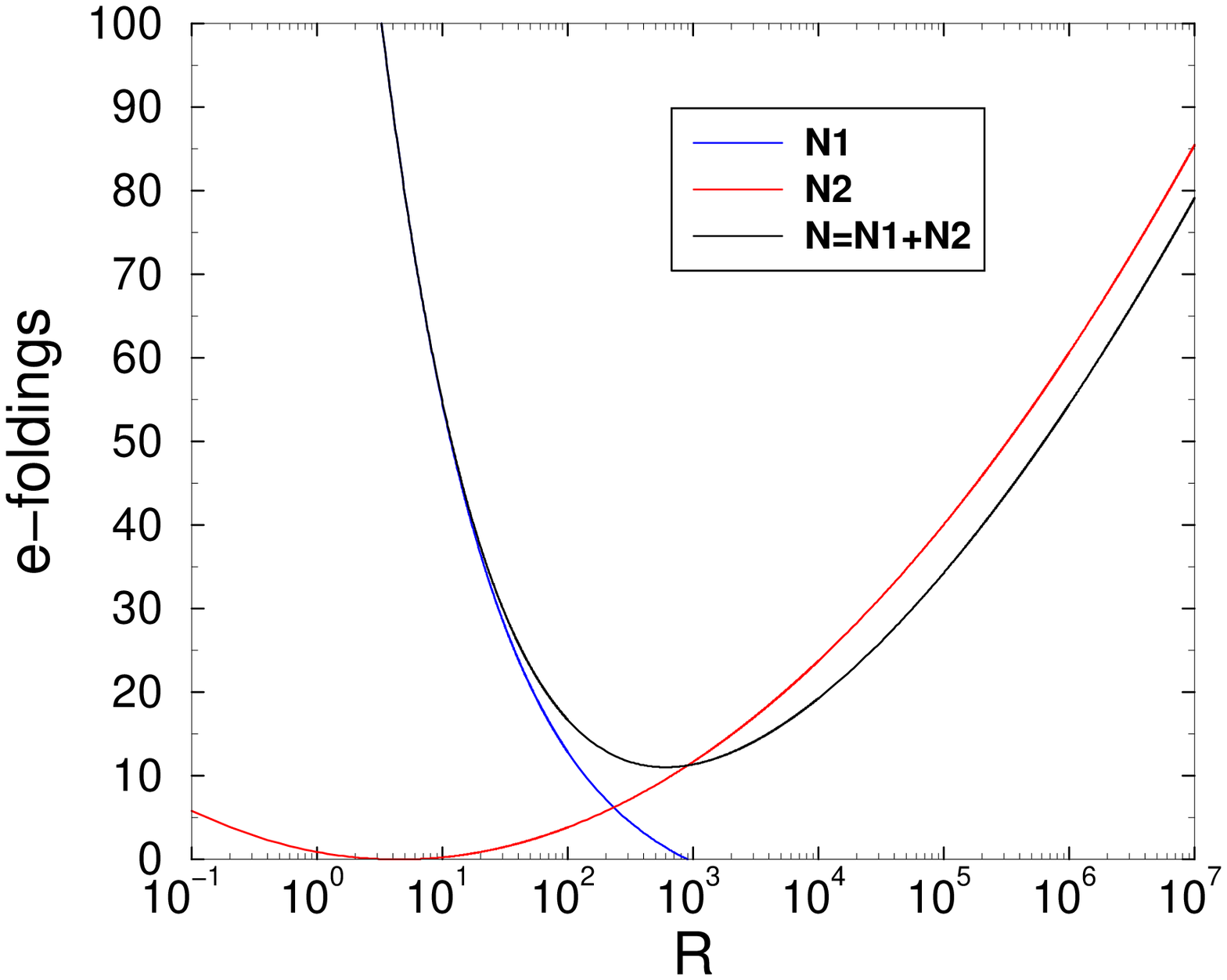}
\end{center}
\caption{\small No. of e-foldings vs. ${\cal R}$}
\label{figps6}
\end{figure}

Next let us try to estimate the number of e-foldings. For this we have to compute the slow roll parameters 
\be
\epsilon_H \equiv 3\frac{\dot{\te}^2}{2V+\dot{\te}^2}\ll 1
\ee 
and
\be
\eta_H\equiv -\frac{\ddot{\te}}{H\dot{\te}}\ll 1
\ee 
First observe that 
\be
V(\te)= V_{\mt{min}}+\4e^{-2v}(\ti{V}(\te)+V_6(v_{\mt{min}},\cd)e^{-2\de'\te})=\4e^{-2v_{\mt{min}}}(2+\ti{V}(\te)+V_6e^{-2\de'\te})
\label{5tpot}
\ee
where we have reintroduced the $X_6$ potential as it  plays an important role away from $\te=0$. $V(\te)$ looks like a Mexican hat. Now inflation can occur as $\te$ rolls down from the maxima to the second minimum along which the potential can be approximated by a cubic potential, the derivative of which has to vanish at the maximum ($-\te_x$) as well as the minimum ($-\te_m$):
\be
V'(\te)=c(\te+\te_x)(\te+\te_m)
\ee
In terms of 
\be
\zeta=-(\te+\te_m)
\ee
\be
V(\ze)=V_m+c(\3\ze^3+\2\Delta\ze^2)
\label{5zpot}
\ee
where 
\be
V_m\equiv V(-\te_m)\mx{ and }\Delta=\te_m-\te_x
\ee
With (\ref{5zpot}) one can compute the total number of e-foldings as
$$
{\cal N}=\frac{V_m}{c\Delta}\left(\ln\frac{\ze_i}{\ze_e}+\ln\frac{\Delta-\ze_i}{\Delta-\ze_e}\right)+\frac{\Delta^2}{6}\left(\ln\frac{\Delta-\ze_i}{\Delta-\ze_e}\right)+\frac{\ze_i-\ze_e}{6}\left(\ze_i+\ze_e-\Delta\right)
$$
\be
\equiv {\cal N}_1+{\cal N}_2+{\cal N}_3
\ee
 $-\ze_i\approx -\Delta$ is where the inflation starts while $-\ze_e\approx 0$ is where the inflation ends. Notice ${\cal N}_3$ is  small and the main contributions to the e-foldings come from ${\cal N}_1$ and ${\cal N}_2$. If we do not  assume any  fine tuning of the initial conditions the logarithms are $\approx$2 so that 
\be
{\cal N}\approx 4\frac{V_m}{c\Delta}+\frac{\Delta^2}{3}
\ee
Now all we have to do is to identify $V_m,c,\te_m$ and $\te_x$ from the actual potential (\ref{5tpot}). Note that $V_6$ is the last to become active, so that $\te_x$ is still given to a good approximation by 
 $\ti{V}(\te)$ and one knows this value exactly \cite{t1,t2}
\be
e^{2\te_x}=\frac{2\Dd-\Dz}{2\Dd+\Dz}
\label{5squashed}
\ee
Next, we realize that the second minimum is produced essentially because of the competition between $X_3$ and $X_6$ so that using (\ref{5vmin}) and (\ref{5tpot}) we get
\be
\te_m=\frac{\Ds}{2\Dz(\Dd-1)}\ln{\cal R}\ ; \ {\cal R}=\frac{k}{V_6}
\ee
and
\be
V_m=\4e^{-2v_{\mt{min}}}\left(\Ds+2-k(\Dd-1){\cal R}^{1/(\Dd-1)}\right)
\ee
${\cal R}$ is basically like a measure of the ratio between $V_{\mt{curv}}$ and $V_{\mt{flux},H}$. 
Further one can compute $V''$ exactly at this minimum
\be
\tilde{V}''_{m}=e^{-2v_{\mt{min}}}k\frac{\Dz(\Dd-1)}{\Ds}{\cal R}^{1/(\Dd-1)}=c\Delta
\label{5vpp}
\ee
which gives us $c$. Using (\ref{5squashed}-\ref{5vpp}) then, one obtains an expression for ${\cal N}$ in terms of the parameter ${\cal R}$, a plot for which looks like fig.~\ref{figps6}. First, observe that even the minimum e-folding that one gets is around 10 which is already of the right order that is required to explain the cosmological problems \cite{cosmos}. Second, note that one can get large number of e-foldings in two regions of small and large ${\cal R}$ although caution should be made in the small ${\cal R}$ region as after a point the second minimum is above the first symmetric minimum and the analysis will break down. 
\vskip 0.1cm
The case of small ${\cal R}$ inflation when most of the e-foldings come from ${\cal N}_1$ signals a hybrid inflationary model \cite{hybrid}. Note here $V_m\neq 0$ and therefore inflation does not end through $\te$ oscillation but rather as the particle approaches the minimum it starts to  roll off in another direction. In the context of our model a natural choice would be the volume $v$ direction or perhaps a second squashing mode which is responsible for later  quintessence! For large  ${\cal R}$ inflation, most of the e-foldings come from ${\cal N}_2$ with the inflaton eventually settling down at the minimum (L-3). In this case it is clear that we cannot solve the cosmological constant problem. ${\cal R}$ has to be fine tuned to make $V_m=0$. In general one has to add a four dimensional cosmological constant by hand to make it vanish. 
\vs
{\bf Constraints from Quintessence:} From the table we see that there are really two possible late time attractor solutions, the $X_4$-region and the $X_{36}$ trough. In either case the particle moves along a constant slope $m_q$ given by
\be
m_q= 0\equiv m_{\mt{L1}}\mx{ for } X_4\mx{-region}
\ee
and
\be
m_q= -\frac{1-\al}{1-\nu}\equiv m_{\mt{L2}}\mx{ for } X_{36}\mx{-trough}
\label{5l2}
\ee
 (\ref{5l2}) follows by applying (\ref{4combv}). During this evolution, the combination of fields 
\be
Q_{\perp}\equiv \te-m_q v
\label{qperp}
\ee
remains ``frozen'' and the particle  rolls along the orthogonal direction $Q$ effectively  along an  exponential potential 
\be
V_{\mt{eff}}\approx \la_q  e^{ -\al_q Q}
\ee
  where the ``exponent'' $\al_q$ is given by
\be
\al_{\mt{L1}}= \frac{2\al}{K_v} = \sqrt{\frac{2\Ds}{\Ds+2}}\mx{ for } X_4\mx{-region}
\label{alphal1}
\ee
and
\be
\al_{\mt{L2}}= \al_{36} = \sqrt{\frac{2\Dz}{\Dz+2}} \mx{ for } X_{36}\mx{-trough}
\label{alphal2}
\ee
It is now clear how a modest rolling of the quintessence field can produce the famed 120 orders of mismatch between Planck mass and the cosmological energy density \cite{cosmos}
\be
\frac{<\la>_{\mt{cos}}}{M_{\mt{p}}^4}\sim 10^{-120}\Ra V_{\mt{eff}}\sim 10^{-120}
\label{5cosm}
\ee
From inspection of (\ref{alphal1},\ref{alphal2}), it is clear that $\al_q$ is smallest  when 
$\Dz=1\Ra \al_q\sim \sqrt{2/3}$ 
while it is always less than $\sqrt{2}$ so that
\be
\sqrt{2}>\al_q\geq \sqrt{2/3}\sim 0.8
\ee
To generate the hierarchy (\ref{5cosm}) then, we just need a rolling of $Q$ from say $Q_0$ to $Q_c$ given by
$$\Delta Q=|Q_c-Q_0|\sim \frac{120}{\al_q}$$
\be
\Ra 150\geq\Delta Q>85
\ee

Let us look at the bounds on equation of state parameter $\om$ coming from observational cosmology \cite{cosmos}:
\be
\om<-0.78
\label{5omega}
\ee
Now, for  exponential potentials the exact late time attractor solutions are known \cite{liddle}. For $\al_q<\sqrt{2}$ this corresponds to a power-law phase of acceleration
\be
a(t)\sim t^{\frac{2}{\al_q^2}}
\label{5evolution}
\ee
with an equation of state parameter $\om_q$ given by
\be
\om_q=-1+\3\al_q^2
\label{5omegaq}
\ee 
Then, compatibility with (\ref{5omega}) seems to tell us 
\be
\al_q\leq 0.8
\ee
which is just  possible when $\Dz=1$ (\ref{5omegaq}). Unfortunately for compact  group manifolds $\Dz=1$ is not possible although for coset space reductions\footnote{ An interesting  example will be squashing  a 2-sphere in a 6-dimensional SUGRA context, which has become specially popular as a brane world model \cite{cliff}.} and non-compact internal manifolds (in the brane-world scenario \cite{dvali}) it is possible to have $\Dz=1$. We find that 
$$\Dz=2\Ra \al_q=1\mx{ while }\Dz,\ \Ds=3\Ra \al_q\sim 1.1$$
These clearly  does not satisfy the current bound (\ref{5omega}). Fortunately this is not the end of the story.  Recently it was discovered that although the late time attractor solution for an exponential potential satisfies (\ref{5omegaq}), before reaching the attractor solution there is a transient phase of acceleration where the equation of state varies \cite{transient}. In particular it was shown in \cite{transient} that this transient phase of acceleration can explain the current phase of acceleration with $\om$ close to -1, even though $\om_q$ is not. Taking this point of view let us now investigate the further constraints coming from violation of equivalence principle and time variation of fine structure constant experiments.

Imposing (\ref{qperp}) one obtains  an effective Lagrangian  which now depends on a single scalar field $Q$:
\be
S_{\mt{eff}}=\int\ d^4x \left[\frac{R}{2}+{\cal L}_{\mt{scalar}}+{\cal L}_{\mt{rad}}\right]
\ee
with
\be
{\cal L}_{\mt{scalar}}=-\frac{1}{2}(\p Q)^2-\la_q e^{-\al_q Q}
\ee
and
\be
{\cal L}_{\mt{rad}}=-\frac{1}{16\pi\al_0}  e^{-\s Q}F^2
\ee
The constant parameters $\al_{\mt{f}}$ and $\la_q$  depend on $Q_{\perp}$. We have used (\ref{qperp}) and normalized $Q$ so that the kinetic terms are canonical
\be
Q= Kv\ ;\ K=\sqrt{K_v^2+m_q^2K_{\te}^2}
\ee
 We now investigate the bounds on the coupling exponent of $Q$ to the electro-magnetic $F^2$ term. Note that in our model for radiation in the bulk, the fine structure constant is time dependent and is given by
\be
\al_{\mt{f}}=\al_0e^{\s Q}
\ee
Then
\be
|\frac{\dot{\al_{\mt{f}}}}{\al_{\mt{f}}}|=|\s|\dot{Q}<10^{-15}yr^{-1}
\ee
where the bound comes from atomic clock experiments \cite{alpha}. For  solutions (\ref{5evolution}) it is known that the kinetic energy is a fraction of the scalar potential energy which is identified with the effective cosmological constant. This gives us a handle to estimate the $Q$ variation
$$\mx{K. E.}=\ti{\om}(\mx{P. E.})\ ;\ \ti{\om}=\frac{1+\om_q}{1-\om_q}$$
Recent CMB data \cite{Riess} seem to suggest $\om_q\sim -1.02\pm 0.1$ using which we find
\be
\Ra \dot{Q}\approx 10^{-13}-10^{-14} yr^{-1}
\ee
This implies a bound on $\s$
\be
|\s|< 10^{-1}- 10^{-2}
\label{5bound}
\ee
Let us compute $\s$ for L-1 and L-2 late time solutions. For bulk radiation from (\ref{3bulk}) one gets in general
\be
\s=\frac{2[1+(1-\nu)m_q]}{K}
\label{5alphabound}
\ee
In particular when $m_{\mt{q}}=-1/(1-\nu)$, $\s=0$ and the bound is obviously satisfied. From fig.~\ref{figps5}, it is clear that the L-2 slope is closer to the $\s=0$ line as compared to the L-1 slope and hence the situation with the bounds \ref{5alphabound} is improved. However it is not close enough. 
Substituting $m_{\mt{L1}}$ and $m_{\mt{L2}}$ in (\ref{5alphabound}) we find
\be
\s_{\mt{L1}}=K_v>\sqrt{2} \mx{ and }\sqrt{\frac{3}{2}}<\s_{\mt{L2}}=\al_{\mt{L2}}<\sqrt{2}
\ee 
respectively. Thus unless the $\om_q$ parameter is significantly closer to -1, $\s_{\mt{L1}}$  cannot satisfy (\ref{5bound}) while $\s_{\mt{L2}}$'s case is marginal. For brane radiation no such problem arises as it is not coupled to the moduli fields. This finding however is consistent with the earlier realization that perhaps our current universe is in a transient quintessence phase rather than having already reached the late time attractor solution.  

Finally we come to the most crucial  observational bound concerning  variation of $\al_{\mt{f}}$ on cosmic scales. Assuming  $\frac{\Delta\al_{\mt{f}}}{\al_{\mt{f}}}$ to be small one finds
\be
\label{variation}
\frac{\Delta\al_{\mt{f}}}{\al_{\mt{f}}}\approx -\s\Delta Q\sim -\s\dot{Q}\Delta t
\ee
where $\Delta t$ is say the time elapsed since BBN. For BBN it is known that $\frac{\Delta\al_{\mt{f}}}{\al_{\mt{f}}}< 10^{-2}$. On the other hand we have seen that typically to be consistent with quintessence cosmology $|\Delta Q|\sim 100$. Thus the first part of the equality gives a very strong bound on $\s<10^{-3}-10^{-4}$. Not only do our late time $\s$'s violate this bound but also it is hard to see how transient quintessence can explain such small $\s$'s without fine-tuning.  However, this has two possible resolutions. Either (a) one has to invent a mechanism such that the quintessence trajectory has a late time attractor slope very close to (or possibly equal to) $ -1/(1-\nu)$  when   $\s=0$. This may be possible to achieve by introducing other flux terms or considering potentials coming from brane gas \cite{vafa} etc.  Or (b) if $|\Delta Q|$ is small! How can this be? Well, all we need is for $\Delta\al_{\mt{BBN}}$ to be small. If we look at the second part of the equation (\ref{variation}), then substituting the typical age of the universe since BBN, $\Delta t\sim 10^{10}yr$, we find  $|\Delta Q|\sim 10^{-3}$! Then  even for $\s\sim 1$,  $\frac{\Delta\al_{\mt{f}}}{\al_{\mt{f}}}$ is well within the observational bounds. Of course this is only an approximate argument and $\dot{Q}$ need not be this small through out  but (b) does seem to offer a viable alternative. What this means of course is that most of the hierarchy in the potential has to come from early (pre-big bang) evolution of $Q$ during inflation, reheating etc. Crudely, the evolution of  the particle can be represented as in fig.~\ref{figps5} where it starts out with an initial pre-BBN slope but then slowly changes (or is still changing) to the late time quintessence slope.

It may now seem that we have solved the problem posed by (\ref{variation}), but   in the bulk case not quite! The reason is that in our potential (\ref{3potential}) $X_3$ is identical to the radiation coupling exponent (\ref{3bulk}) modulo a $-$ sign. For the current value of $\al_{\mt{f}}$ this means that   $X_3$  provides a large contribution to the cosmological constant and  neither resolutions (a,b) can work without resorting to fine tuning of the parameters in order to cancel the $X_3$ contributions,  but  this is precisely what quintessence models aim to avoid! The only way to avoid these fine tunings is to ensure $k=0$ when  the $X_3$ potential term is absent. Remarkably, one finds \cite{t2} that  $k$, which is determined by group theory, vanishes   whenever $H$ is a $U(1)$ or a product of  $U(1)$'s. This is nice if one eventually wants to make a connection to symmetry breaking in Standard Model. 

Finally, one can look at the  various null experiments on fifth force \cite{review}. In general the quintessence field will also couple to matter
\be
S_{\mt{matter}}=\int d^4x\ \sqrt{-g}\bar{\chi}(\nabla{\hskip-3.0mm}/+ime^{\mu Q})\chi
\ee
 and this will  lead to violation of equivalence principle \cite{review}. However typically bounds obtained from these experiments come in the combination of $\mu^2$ and $\mu\s$:
 \be
 \mu\s<10^{-10}
 \ee
 and this only tells us that for the values of $\s$ in our model
 \be
 \mu<10^{-9}
 \ee
 and we leave the study of fermionic couplings to future research.
\section{{\bf SUMMARY AND FUTURE RESEARCH}} 

To summarize, we have derived potentials for the volume and the squashing mode in a typical SUGRA model. The potential gets contributions from the internal curvature, the flux, and the higher dimensional cosmological constant. It turns out generically to be a sum of six exponentials with rich features which we explored to understand the corresponding cosmological phases. We find that at late times the moduli could either be stabilized around a minimum (L-3), or one of them (a linear combination of shape and size) could be frozen while the other evolves towards $\pm$infinity (L-1,2). In this latter case one obtains quintessence as the external universe undergoes accelerated expansion for these late time attractor solutions. Prior to reaching these late time solutions, the universe in general undergoes radiation-matter domination as the moduli potential is steep (dominating effective exponents being larger than $\sqrt{2}$). Thus, for a class of initial conditions our potential predicts the history of our universe to be radiation-matter $\ra$ quintessence (acceleration). On top of these phases our model can also accommodate an early phase of acceleration or inflation if and when the internal manifold makes a transition from the symmetric (unsquashed) minimum towards the asymmetric (squashed) minimum. In the original Kaluza-Klein picture where the four dimensional gauge fields originate from the higher dimensional (bulk) graviton, this phenomena also corresponds to a gauge symmetry breaking mechanism. For brane gauge fields no such interpretation is possible. 

To understand how these various phases of cosmology emerge, both qualitatively and quantitatively, we developed techniques to determine the moduli evolution under the influence of exponential potentials. These techniques work for any number of exponentials with two fields and perhaps can be generalized to more than two fields. Hence, one could add more fluxes or introduce brane-gas etc, and study their effect on the evolution/stabilization. Finally, we should note that although we specialized to group internal manifolds, most of the results (including the exponents) only depend on the various dimensions in the problem and hence are completely general to compact internal manifolds. Some specific details may change because the coefficient in front of the exponentials can vary and it would be interesting to look at coset spaces, specially those whose isometry group is similar to the gauge group of the Standard model or GUTs. 

The analysis and the results so far are encouraging, but several problems still remain before we can envisage realistic particle physics and cosmology. First let us focus on the quintessence aspect. From our study of several constraints coming from variation of fine structure constant and the fifth force, several fine tuning issues persist with regards to the radiation coupling exponent $\s$. This is enhanced if one assumes universal coupling or Brans-Dicke type theory. There are essentially three ways to avoid these constraints: (i) one invents a mechanism so that the linear combination that is stabilized corresponds precisely to $\s=0$. Although we came close, we didn't really succeed. There is still scope of adding other fluxes and/or introducing brane gas etc. which can do the trick but again this will only work for certain specific dimensions! (ii) one finds a mechanism like chameleon mechanism to ameliorate bounds on $\s$. (iii) show that most of the scalar evolution that occurs precedes BBN, while since (or sometime before it) BBN the scalars have really slowed down so that they are effectively frozen. 

Next let us look at the inflationary scenario. The main problem associated with the inflationary scenario in our model is to have a graceful exit. If the second minimum is really deep then inflation can end in the usual way with oscillations reheating the universe. However, in this case one cannot address the issue of the small cosmological constant as one has to fine tune the parameters to ensure that the potential energy at the minimum is zero (or close to it). The  other choice is to have a quintessential-inflation type scenario where the quintessence field and the inflaton are nevertheless separate fields. As the inflaton approaches the minimum it veers off in another (steeper) direction as in hybrid inflationary models except that this new direction (or field) plays the dual role of a quintessence field later on! This is a nice picture particularly because our model naturally contains two fields (for example $\te$ could be the inflaton, while $v$ plays the dual role) except that it doesn't quite work. First, notice that quintessence has a chance to work only if $k=0$, on the other hand we found a second minimum (and flattish potentials) only when $k>0$. Second, if we assume $k>0$ for successful inflation, the particle veers on to the $X_{36}$ trough which also supports acceleration. Similarly, if we assume $k=0$ the particle veers off along $v$ direction under the influence of $X_4$ exponent, again signalling an accelerated phase. Hence ending inflation in these scenarios is a nontrivial task. We believe one can address this issue in two different ways. (i) As pointed out in \cite{transient} the particle does not immediately fall into the late time attractor solutions ($X_{36}$ or $X_4$) but rather there is a transitional phase of radiation-matter  domination. One can investigate whether such a phase can account for the radiation era in our universe. (ii) A more attractive scenario would be to invoke double squashing! Two successive transitions, one with $k>0$ supporting inflation, followed by a $k=0$ transition taking care of the graceful exit problem along with later quintessence.

This second scenario becomes even more appealing in the light of our common understanding of particle physics. Our universe is supposed to have undergone two gauge symmetry breakings -  first, a symmetry breaking of a GUT gauge group to the Standard Model gauge group and second the Electro-weak symmetry breaking. Note for the second transition $k=0$, a $U(1)$ is involved which is precisely what we need for the quintessence to work! The challenge of course now is to come up with the right internal manifold. If one wants to accomodate all the symmetries within the seven extra dimensions one has to look into coset spaces whose consistent truncations are slightly more difficult to study. On the other hand, even as a toy exercise, if one wishes to first develop a group manifold  model, one has to generalize our constructions. This is because the Standard Model gauge group is not simple, not even semi-simple. This would involve including more squashing fields corresponding to the different simple components in order to have a consistent truncation and we leave this exercise for future research. 

To conclude several challenges still remain but perhaps we are just beginning to understand the implications of extra dimensional dynamics. 
\vspace{5mm}\\
{\large {\bf Acknowledgments:}} The authors would like to thank Anupam Mazumdar for some useful discussions and suggestions. \vskip 0.1cm

This work is supported in part by
the Natural Sciences and Engineering Research Council of Canada and in
part by the Fonds Nature et Technologies of Quebec.


\begin{thebibliography}{99}

\bibitem{kachru}  S.B. Giddings, S. Kachru and J. Polchinski {\it Phys.Rev.} {\bf D66} (2002) 106006, hep-th/0105097; S. Kachru, M. Schulz and S. Trivedi {\it JHEP} {\bf 0310} (2003) 007, hep-th/0201028
\bibitem{kaluza} T. Applequist, A. Chodos and P.G.O. Freund {\it Modern Kaluza-Klein Theories} Addison-Wesley Publishing Co. Inc. (1987); M.J. Duff, C.N. Pope and B.E.W. Nilsson {\it Phys. Rep.} {\bf 130} (1986) 1
\bibitem{duff}  M.J. Duff, C.N. Pope, N.P. Warner and B.E.W. Nilsson {\it Phys. Lett.} {\bf 149 B} (1984) 90;  M.J. Duff and  C.N. Pope {\it Nucl. Phys.} {\bf B 255} (1985) 355
\bibitem{t2} T. Biswas and P. Jaikumar, {\it Phys. Rev.} {\bf D 70} (2004) 044011, hep-th/0310172
\bibitem{kachru2}  E. Cremmer and J. Scherk,  Nucl. Phys. {\bf B108}, 409 (1976); R. Sundrum, Phys. Rev. {\bf D59}, 085010 (1999); N. Arkani Hamed, S. Dimopoulos and J. March-Russell, Phys. Rev. {\bf D63}, 064020 (2001); S. Kachru, R. Kallosh, A. Linde and S.P. Trivedi {\it Phys.Rev.} {\bf D 68} (2003) 046005, hep-th/0301240; U.Guenther, P.Moniz and A. Zhuk, {\it Phys. Rev.} {\bf D 68},(2003) 044010, hep-th/0303023
\bibitem{ratra} P.J.E. Peebles  and B. Ratra {\it Astrophysics J.} {\bf L 17 } (1988) 352; R.R. Caldwell, R. Dave and P.J. Steinhardt {\it Phys. Rev. Lett.} {\bf 80} (1988) 1582
\bibitem{cosmos} N.A. Bahcall, J.P. Ostriher, S. Perlmutter and P.J. Steinhardt {\it Science} {\bf 284 } (1999) 1481, astro-ph/9906463; A.G. Reiss et. al.  {\it AJ} {\bf 116} (1998) 1009, astro-ph/9805201; S. Perlmutter  et. al. {\it AJ} {\bf 517 } (1999) 565, astro-ph/9812133
\bibitem{cliff} Y. Aghababaie, C.P. Burgess, S.L. Parameswaran, F. Quevedo  {\it JHEP} {\bf  0303} (2003) 032, hep-th/0210233;  {\it Nucl.Phys.} {\bf B680} (2004) 389, hep-th/0304256; Y. Aghababaie, C.P. Burgess, J.M. Cline, H. Firouzjahi, S.L. Parameswaran, F. Quevedo, G. Tasinato and I. Zavala {\it JHEP} {\bf  0309} (2003) 037, hep-th/0308064
\bibitem{alpha} See for e.g. L. Anchordoqui and H. Goldberg, {\it Phys. Rev.} {\bf D68} (2003) 083513, and references there in.
\bibitem{review} Clifford M. Will {\it Living Rev.Rel.} {\bf 4 } (2001) 4, gr-qc/0103036, and references there in
\bibitem{vafa} R. Brandenberger and C. Vafa {\it Nucl. Phys.} {\bf B 316 } (1989) 391; S. Alexander, R.H. Brandenberger and D. Easson {\it Phys. rev.} {\bf D 62 } (2000) 103509, hep-th/0005212; T. Battefeld and S. Watson {\it JCAP} {\bf 0406} (2004) 001,  hep-th/0403075
\bibitem{justin} Justin Khoury and Amanda Weltman {\it Phys.Rev.} {\bf D69} (2004) 044026, astro-ph/0309411
\bibitem{frozen}   A. Mazumdar  {\it Phys. Lett.} {\bf B 469}, 55 (1999); hep-ph/9902381
\bibitem{guth}  A.A. Starobinsky {\it Phys. Lett.} {\bf 91 B} (1980) 99; A.H. Guth  {\it Phys. Rev. } {\bf D 23} (1981) 347; A. Linde {\it Phys. Lett.} {\bf 108 B} (1982); {\bf 129 B} (1983) 177; A. Albrecht and P.J. Steinhardt  {\it Phys. Rev. Lett.} {\bf 48} (1982) 1220
\bibitem{linde}  For reviews see for e.g. A. Linde, Particle Physics and Inflationary Cosmology {\it harwood academic publishers} (1990);  F. Englert {\it Phys. Lett.} {\bf 119 B} (1982) 339; A.R. Liddle, astro-ph/9901124; S. Watson, astro-ph/0005003
\bibitem{t1} T. Biswas {\it JHEP} {\bf 0302} (2003) 054, hep-th/0210273 
\bibitem{riet} A. Collinucci, M. Nielsen, T. Van Riet,  hep-th/0407047
\bibitem{liddle} D. Wands, E.J. copeland and A.R. Liddle {\it Ann. of N.Y. Acad. Sci.} {\bf 688 } (1993) 647; {\it Phys. Rev.} {\bf D 57} (1998) 4686, gr-qc/9711068;  P.G. Ferreira and M.Joyce {\it Phys. Rev. Lett.} {\bf D 79} (1997) 4740, astro-ph/9707286; T. Barreiro, E.J. copeland and N.J. Nunes {\it Phys. Rev.} {\bf D 61 } (2000) 127301, astro-ph/9910214
\bibitem{anupam}  A. R. Liddle, A. Mazumdar and F. E. Schunck, {\it Phys. Rev.} {\bf D58}, 061301 (1998); E.J. Copeland, A. Mazumdar and N.J. Nunes, {\it Phys. Rev.} {\bf D60}, 083506 (1999)Justin Khoury and Amanda Weltman {\it Phys.Rev.} {\bf D69} (2004) 044026, astro-ph/0309411
\bibitem{wohlfarth}  M.N.R. Wohlfarth {\it Phys.Rev.} {\bf D 69} (2004) 066002, hep-th/0307179
\bibitem{kerner} R. Kerner {\it Ann. Inst. H. Poincare} {\bf 9} (1968) 143
\bibitem{pope}  M. Bremer, M.J. Duff, H. Lu, C.N. Pope and K.S. Stelle {\it Nucl. Phys.} {\bf B 543} (1999) 321, hep-th/9807051
\bibitem{ohta} N. Ohta, {\it Phys. Rev. Lett.} {\bf 91}, (2003) 061303,   hep-th/0303238; {\it Prog. Theor. Phys.} {\bf 110}, (2003) 269, hep-th/0304172; C.-M. Chen, P.-M. Ho, I. P. Neupane, N. Ohta and J. E. Wang, {\it JHEP} {\bf 0310},  (2003) 058,  hep-th/0306291; S. Roy  {\it Phys.Lett.} {\bf B 567} (2003) 322, hep-th/0304084
\bibitem{transient} E. Bergshoeff, A. Collinucci, U. Gran, M. Nielsen, D. Roest {\it Class. Quant. Grav.} {\bf 21} (2004) 1947, hep-th/0312102
\bibitem{Sundrum} L. Randall and R. Sundrum {\it Phys. Rev. Lett.} {\bf 83} (1999) 3370, hep-th/9906064; 4690, hep-ph/9905221
\bibitem{hybrid} A.D. Linde,  {\it Phys. Lett.} {\bf B 259}, 38 (1991);  Phys. Rev. {\bf D49}, 749 (1994)
\bibitem{peebles} P.J.E. Peebles and A. Vilenkin {\it Phys. Rev.} {\bf D 59} (1999) 063505, astro-ph/9810509
\bibitem{dvali} G.R. Dvali, G. Gabadadze and M. Porrati {\it Phys.Lett.} {\bf B 485 } (2000) 208
\bibitem{Riess} A.G. Riess et al, {\it ApJ} in press, to appear June 2004,
 astro-ph/0402512
\end{thebibliography}
\end{document}